# UV-cured PVA-collagen co-network with staining-free photodynamic activity against chronic wound bacteria

Nikitia Mexia,[1] Charles Brooker,[1,2] Giuseppe Tronci[1,2*]

[1]Clothworkers Centre for Textile Materials Innovation for Healthcare (CCTMIH), School of Design, University of Leeds, Leeds, LS2 9JT, United Kingdom

[2]School of Dentistry, St. James's University Hospital, University of Leeds, LS9 7TF, United Kingdom

## Abstract

Chronic wounds affect 1-2% of the global population and represent an unsustainable healthcare, financial and economic burden. Delayed healing is largely associated with bacterial infection and inadequate moisture balance, both of which remain difficult to control due to increasing antimicrobial resistance and the limited exudate management capacity and mechanical performance of current dressings. To address these challenges, we introduce a staining-free photodynamic material with work-light-triggered antibacterial activity, significant water-induced swelling (up to more than nine-fold increase in dry weight and volume) and tunable wet state compression modulus ($E_c$: 76±34−289±94 kPa). The material was prepared by UV co-curing of an aqueous solution containing a bespoke poly(vinyl alcohol) (PVA) derivative functionalised with both phenothiazine and methacrylate residues, 4-vinylbenzyl-functionalised type 1 rat tail collagen, and a water-soluble photoinitiator. Fourier Transform Infrared (FTIR) spectroscopy and Differential Scanning Calorimetry (DSC) indicated the presence of characteristic chemical and thermal features of both components within the co-networks. 96-hour co-network incubation *in vitro* (1X phosphate buffered saline, pH 7.4, 37°C) confirmed the significant retention of the phenothiazine dye (> 93 wt.%), i.e. toluidine blue (TB), suppressed TB-induced colouration of the surrounding medium, and an average mass retention higher than 98 wt.%, in contrast to an interpenetrating network control prepared with a PVA derivative lacking methacrylate residues. 2',7'-Dichlorofluorescein assay demonstrated the generation of reactive oxygen species (ROS) after 10-minute exposure of the UV-cured co-networks to work light (~15 mW $cm^{-2}$), whereas minimal ROS-associated fluorescence was observed under corresponding dark conditions. Exposure of the samples to work light for 10 min resulted in at least 1 $log_{10}$ reduction in bacterial viability against both gram-positive *S. aureus* and gram-negative *P. aeruginosa*, while no bacterial reduction was detected in dark-control conditions. Importantly, the metabolic activity of L929 fibroblasts remained above 90% following illumination. The antibacterial properties, staining-free photodynamic behaviour, water-induced swelling, and wet-state mechanical properties achieved via network integration with functionalised collagen, make this material appealing for chronic wound healing applications and for on-demand delivery of localised therapies.

**Highlights**

- UV-cured co-network from TB-conjugated PVA methacrylate and photoactive collagen
- Significant water-induced co-network swelling (> 9-fold increase in weight /volume)
- ROS production after 10-minute irradiation under work light
- Photodynamic inactivation of both Gram-positive and Gram-negative bacteria
- Preserved metabolic activity of L929 murine fibroblasts

## 1. Introduction

The skin is the largest organ of human body and consists of three layers (epidermis, dermis and hypodermis), forming a barrier that protects the body against many exterior hazards, including chemicals and microorganisms.

* Corresponding author at: Level 7 Wellcome Trust Brenner Building, St. James's University Hospital, Leeds LS9 7TF, United Kingdom. E-mail address: g.tronci@leeds.ac.uk (G. Tronci)

Various events including injuries, burns, surgery and illness, can disrupt the structural integrity of the skin, leading to the formation of wounds and increasing susceptibility to infection [1-6].

Wound healing is a complex process that includes phases of haemostasis, inflammation, proliferation and remodelling [7]. Depending on the severity of the injury, a wound can be classified as either acute or chronic. In the case of acute wounds, the healing process occurs within 8-12 weeks without the need for clinical intervention [4,5,8]. On the other hand, the healing process of chronic wounds is complex and takes significantly longer time, sometimes requiring years or even decades [5,8,9]. A wound is usually classified as chronic if it exhibits minimal indications of improvement beyond three months [10], as observed in diabetic foot ulcers, burns, ulcerative colitis, and pressure ulcers [5,7]. Risk factors associated with the development of chronic wounds include age, immunological status, malnutrition, infection, tobacco use, underlying disorders, drugs, exposure to radiation, and chemotherapy [6,11]. Chronic wounds are often accompanied by chronic inflammation and chronic pain, while prolonged exposure to the external environment make them susceptible to bacterial infections, such as those triggered by *P. aeruginosa*, *S. aureus*, *E. coli* and *B. subtilis*. These detrimental biochemical shifts and current challenges in point-of-care diagnosis often lead to preventative overuse of anti-inflammatory drugs and antibiotics, with bacterial resistance being a fundamental challenge [4,6,8,9]. Overall, chronic wounds substantially influence the quality of life of affected people and have a substantial economic impact on health systems worldwide. It is estimated that about 1% of the adult population worldwide suffers from a chronic wound, so that wound management is estimated to account for 3% of the total medical expenditure every year. In the United Kingdom, chronic wounds cost the National Health Service (NHS) £5bn per annum and the projected expenditure for venous leg ulcers in the United States of America is expected to total up to USD 3.5 billion a year [6,9,12].

Successful chronic wound therapies aim to regulate the wound environment to support a pro-healing milieu. Given the multiple factors associated with wound chronicity, a concept that is often used in clinical practice is M.O.I.S.T., where M stands for moisture balance, O for oxygen balance, I for infection control, S for supporting strategies, and T for tissue management [12]. A variety of dressings have been developed to protect the wound environment from external contamination and moisture loss, though challenges still persist with respect to efficient exudate management *in situ*. Dressings made from natural materials such as cotton have initially been used to manage fluids produced by the wound, given their well-known biocompatibility [6,14]. However, the recovery in this case was often slow so that attention has moved to the use of synthetic biomaterials in wound dressings [8,13], due to their highly modifiable physical attributes and adhesive characteristics. Dressings have been developed in the form of semipermeable membranes, foams, hydrocolloids and hydrogels, with the latter ones attracting significant attention given their porosity, hydrophilicity, and cellular tolerability.[5] Hydrogels exhibit a large water content (>90 wt.%), which can be exploited to ensure a moist environment on the wound surface and to support tissue repair. Other qualities of hydrogels include firm adhesion, shape adaptability and mechanical protection [6,15,16]. Additionally, hydrogels can be customised into porous textile products, whose internal porosity offers advantages towards wound exudate management and mechanical integrity in the wet state [17].

Together with moist wound environment, infection management is also critical given that the prolonged wound exposure to the exogeneous environment favours colonisation with pathogens. Wound bacteria frequently exhibit antibiotic resistance, a phenomenon showing a rising trend worldwide, highlighting the need for new antimicrobial chemistries and biomaterial design approaches. Antibiotic-free antimicrobial activity that can be readily administered on the wound whilst being environmentally friendly, localised and associated with minimal risk of generating resistance and ecotoxicity, is a highly sought-after solution [18]. A promising concept that has emerged in recent decades is antimicrobial photodynamic therapy (aPDT), a treatment that can eliminate microorganisms through activation of oxidative stress in a controlled, on-demand manner, with no reliance on environmentally toxic metals, such as silver [19]. aPDT involves the application of a photosensitiser (PS) and appropriate light excitation in presence of molecular oxygen to generate reactive oxygen species (ROS), which can cause damage to surrounding pathogens. Control of PS type and concentration, and exposure time, can yield high efficiency, while minimising toxicity to endogenous tissues and cells [20,21]. While aPDT can trigger multiple ROS-induced antimicrobial pathways so that resistance risks are minimal, the reduced generation of ROS, and PS-induced staining, in aqueous media pose challenges to therapeutic reproducibility and clinical uptake [19,21,22].

In this work, we report the creation of a UV-cured co-network consisting of functionalised rat tail collagen and a photodynamic and photocurable derivative of PVA. While both polymers have shown applicability in wound dressings [5,23], there remains an unmet need for advanced wound care solutions that achieve antibiotic-free and silver-free antibacterial activity while ensuring effective exudate absorption, mechanical integrity and cellular tolerability in physiological environments [6,24,25]. To address these challenges, we hypothesised that the presence of styrenic and methacrylate functionalities in selected raw materials could be exploited to achieve a UV-induced, chemically cross-linked co-network integrating the hydrophilicity and wound healing capability of the functionalised collagen system and the photodynamic antibacterial activity of PVA [8,26-28]. The conjugation of TB to the PVA backbone enables antibacterial activity following light exposure of the resulting co-network, while the staining and passive release typically associated with the application of freely dispersed PSs are suppressed [18]. Following the principles of conventional PDT, co-network exposure to visible light is expected to trigger excitation of grafted TB residues and reaction with molecular oxygen, ultimately generating ROS-induced inactivation of common chronic wound bacteria.

## 2. Materials and Methods

### *2.1. Chemicals*

PVA ($M_w$: 146,000−186,000 g·mol$^{-1}$, 99% hydrolysed), *p*-Toluenesulfonyl chloride (TsCl), Toluidine Blue *O*, methacrylic anhydride (MA), triethylamine (TEA), 2-Hydroxy-4-(2-hydroxyethoxy)-2-methylpropiophenone (I2959), 4-vinylbenzyl chloride and 2',7'-dichlorodihydrofluorescein diacetate were purchased from Sigma Aldrich. Acetone, ethanol, DMSO and 1,1,1,3,3,3-Hexafluoro-2-propanol (HFIP) were purchased from VWR, Sigma Aldrich, and Fluorochem, respectively. Rat tail collagen was prepared by extraction of collagen from the tails of pet shop-bought rats according to the previously developed protocol [29]. Minimum Essential Medium Eagle, Alpha Modification (alpha-MEM), trypsin, foetal bovine serum (FBS), L-glutamine (Gln), and penicillin–streptomycin (P-S) were purchased from Sigma Aldrich. Phosphate buffered saline (1X PBS) was purchased from Corning Incorporated. Calcein-AM/ethidium homodimer Live/Dead assay and alamarBlue™ Cell Viability Reagent were purchased from Thermo Fisher.

### *2.2. Synthesis of functionalised PVA products*

*Tosylated PVA (PVA-Ts)*. PVA was dissolved in distilled water (3 wt.% PVA) at 90 °C under magnetic stirring. The solution was cooled down to room temperature under stirring and then TsCl was added at a ratio of 50 mol.% with respect to the vinyl alcohol repeat unit, followed by addition of TEA (1 eq. of hydroxyl groups). The solution was heated at 90 °C and the reaction was left under magnetic stirring for 24 h. The mixture was subsequently cooled down to room temperature and added in a 5-fold excess of acetone. Following overnight precipitation, the pellet was recovered after filtration of the supernatant and dried overnight at 37 °C.

*TB-conjugated PVA (PVA-TB).* PVA-Ts was dissolved in distilled water (3 wt.% PVA-Ts) under magnetic stirring at 90 °C. The solution was cooled down to room temperature under stirring, prior to addition of TB (0.5 mol.% of vinyl alcohol repeat unit) and TEA (1 eq. of hydroxyl groups). The solution was heated at 90°C and the reaction was left under magnetic stirring for 48 h. The mixture was again cooled down to room temperature and added in a 5-fold excess of acetone. The product was recovered by filtration, dissolved again in a small amount of distilled water (approx. 1/10 of the original water volume, at 80 °C) and re-added in acetone. This purification procedure was conducted a third time to remove any unreacted dye and the final pellet was collected by filtration and dried overnight at 37 °C.

*TB-conjugated PVA methacrylate (PVA-TB-MA)*. PVA-TB was dissolved in distilled water (3 wt.% PVA-TB) at 80 °C under magnetic stirring and the solution was cooled down to room temperature prior to the addition of MA (50 mol.% of vinyl alcohol repeat unit) and TEA (1 eq. of hydroxyl groups). The mixture was heated at 60 °C and left for 5 h under magnetic stirring, followed by cooling and addition in a 5-fold excess of acetone. The purification of the product was the same as for PVA-TB, the pellet was collected by filtration and dried overnight at 37 °C.

### *2.3. Chemical characterisation*

Attenuated total reflection Fourier transform infrared (ATR–FTIR) spectra were recorded with dry polymer samples using a Spectrum Three FT-IR Spectrometer (PerkinElmer, Waltham, MA, USA) with a Golden Gate

ATR attachment (Specac Ltd., London, UK). Scans were conducted from 4000 to 600 $cm^{-1}$ with 100 repetitions averaged for each spectrum. Resolution and scanning intervals were 4 $cm^{-1}$ and 2 $cm^{-1}$, respectively.
Proton nuclear magnetic resonance ($^1H$ NMR) spectra were recorded (400 MHz, Bruker) by dissolving 8 mg of polymer in 0.8 mL of deuterated Dimethyl Sulfoxide ($d_6$-DMSO, Sigma Aldrich) assisted by ultrasound sonication (Grant Ultrasonic bath, XB2, Grant Instruments (Cambridge) Ltd.).
Absorption spectra of the functionalised products PVA-Ts, PVA-TB and PVA-TB-MA were measured using a microplate reader (Varioskan LUX, Thermo Scientific). Full spectra were recorded for the product solutions in DMSO (10 mg $mL^{-1}$, n =2) and the absorbance values of the DMSO control solution were subtracted from the absorbance values measured for the products. Additionally, full spectra of 12 solutions of TB in DMSO at different concentrations were recorded. Linear fitting ($R^2$ >0.98) of the absorbance values at 640 nm generated a calibration curve (Figure S1, Supp. Inf.), which was used for the quantification of TB coupling in PVA-TB and PVA-TB-MA, respectively.

### *2.4. Preparation of PVA-based films and UV-curing of the polymer samples*

PVA-TB-MA was dissolved (3 wt.%) in HFIP or distilled $H_2O$ solutions containing 1 wt.% of photoinitiator (I2959), with the help of magnetic stirring at 80 °C in the dark. The polymer solution (either 0.5 or 0.2 mL) was cast into silicone moulds (Ø: 15 mm, h: 3 mm) and left to either air dry in a chemical fume hood (for the polymer solutions in HFIP) or in an oven at 60 °C (for the polymer solutions in distilled water) for up to 48 h.
Additionally, photoinitiator-free solutions of PVA-TB-MA (3 wt.%) were prepared in HFIP via magnetic stirring at 90 °C and cast in the same moulds and left to air dry in a chemical fume hood.

### *2.5. UV-curing of PVA samples*

The photoinitiator-supplemented PVA-TB-MA dry samples were UV-cured by irradiation for 15 min on the top and the bottom side respectively (365 nm, Chromato-Vue C-71, Analytik Jena, Upland, CA, USA). After curing, the films were washed with distilled water (3 h 30 min total time with medium change every 30 min, 3 mL per film) and air-dried. The UV-cured samples of PVA-TB-MA were labelled as PHOVA (dry curing).

### *2.6. Preparation of UV-cured PVA-collagen co-networks*

PVA-TB-MA was dissolved (3 wt.%) in an I2959-supplemented solution of distilled water (1 wt.%) under magnetic stirring at 80 °C and equilibrated to room temperature prior to integration with a functionalised collagen solution [17]. Type I rat tail collagen was reacted with 4-vinylbenzyl chloride according to a procedure previously described [30]. The resulting photoactive product revealed a degree of functionalisation of 25±2 mol.% (n=9), which was measured via 2,4,6-trinitrobenzenesulfonic acid (TNBS) colourimetric assay (Figure S2, Supporting Information) and found to be in line with previous reports [29,31]. 4-vinylbenzyl-functionalised rat tail collagen (4VBC) was dissolved (1.2 wt.%) in a solution of 10 mM hydrochloric acid supplemented with 1 wt.% I2959, under magnetic stirring at room temperature in the dark. For the preparation of the collagen homo-networks, the photoactive solution of 4VBC was cast on a 24-well plate (1.02 g per well) and UV-cured as indicated above, producing the samples 4VBC*.
For the preparation of the PVA-collagen co-networks, the previously prepared photoactive solutions of PVA-TB-MA and 4VBC were cast on a 24-well plate to achieve an overall polymer concentration of 2 wt.% (approx. total weight 1.3g per well) and varying molar ratios between MA residues (of PVA) and 4-vinylbenzyl residues (of collagen). The mixtures were stirred to achieve a homogeneous solution prior to UV-irradiation for 25 min on top and bottom side, respectively. Resulting samples were labelled P2VAXX, where the number 2 indicates the total polymer content of the samples (2 wt.%) and XX indicates the molar excess of MA residues relative to 4-vinylbenzyl residues in the corresponding co-network-forming solutions.
UV-cured co-networks were also obtained from blended solutions with an overall polymer concentration of 3 wt.% and labelled as P3VAXX. PVA-TB-MA was dissolved in a solution of 10 mM hydrochloric acid supplemented with 1 wt.% I2959 via magnetic stirring at 80 °C. Following equilibration to room temperature, 4VBC was added under continued magnetic stirring in the dark to achieve an overall polymer concentration of 3 wt.%. The mixture was then cast on a 24-well plate (1.02 g per well) and UV-cured by 25-min irradiation of top and bottom sides. The same method was followed for the preparation of an interpenetrating network control, coded as I3VA35, where PVA-TB, rather than PVA-TB-MA, was used.

Following UV-curing, samples of either 4VBC*, P2VAXX, P3VAXX and I3VA35 were incubated in a solution of 10 mM hydrochloric acid overnight (3 mL/film) and then washed thoroughly according to the following scheme (Table S1, Supp. Inf.): 3.5 h in distilled water (medium change every 30 min), 15 min in 80:20-distilled water-ethanol mixture, 15 min in 60:40-distilled water-ethanol mixture, 15 min in 40:60-distilled water-ethanol mixture, 15 min in 20:80-distilled water-ethanol mixture, 1.5 h in ethanol (medium change every 30 min).

## *2.7. Mechanical tests and thermal analysis*

Compression tests were carried out on water-swollen UV-cured dressing composite films (Ø: 10 mm) at room temperature with a compression rate of 3 mm $min^{-1}$ (BOSE EnduraTEC ELF 3200, EnduraTEC Systems Corporation, Minnetonka, MN, USA). Stress–compression curves were recorded until load limit was reached, using a 10 N load cell. The compression modulus was quantified by linear fitting in the strain range associated with 10–15% of maximal stress. At least three replicates were employed for each composition. Data are presented as mean ±SD.

Tensile tests (INSTRON 5544, 10 N loading cell) were carried out on individual water-swollen stick-form UV-cured co-networks (length: ~7 mm, width: ~3mm) using an elongation rate of 3 mm $min^{-1}$ and a gauge length of 10 mm. Tests were conducted using a 5 N load cell; the resulting stress–strain curves were fitted linearly in the strain range associated with 10-15% of ultimate stress for the quantification of the Young's modulus.

DSC (PerkinElmer DSC 4000, Massachusetts, USA) was performed on dehydrated samples (~0.5 mg). Thermograms were recorded according to the following program: heating from −10 °C to 240 °C at a heating rate of 10 °C $min^{-1}$, stable at 240 ℃ for 5 minutes, cooling to -10 ℃ at a cooling rate of 10 °C $min^{-1}$, stable at -10 ℃ for 5 minutes, heating to 240 ℃ at a heating rate of 10 °C $min^{-1}$. Prior to measurement, the DSC cell was calibrated using indium as the standard under a nitrogen flow rate of 30 $cm^3$ $min^{-1}$.

## *2.8. Gel content and swelling tests*

The gel content (*G*) was measured to confirm UV-induced network formation and quantify the overall portion of water-insoluble sample. Dry UV-cured samples (n =3-4) were weighed ($m_d$) and individually incubated in 3 mL of a 10 mM solution of hydrochloric acid (room temperature, 24 h), prior to air-drying and weighing ($m_1$). *G* was calculated through the equation:

$$G = \frac{m_1}{m_d} \times 100 \qquad (1)$$

Swelling tests were performed on dry UV-cured samples of known mass ($m_d$) via incubation in 3 mL of 1X PBS (pH 7.4, 37 ℃), followed by gravimetric and dimensional analysis. The mass of water-equilibrated samples ($m_s$) was recorded (n =3-4) and the swelling ratio (*SR*) was calculated through the following equation:

$$SR = \frac{m_s - m_d}{m_d} \times 100 \qquad (2)$$

The volumetric swelling (*Q*) was quantified by measuring the diameter and height of the sample discs in the dry state and following the 24-hour incubation in 1X PBS (pH 7.4, 37 ℃). *Q* was calculated through the following equation:

$$Q = \frac{V_s}{V_d} \times 100 \qquad (3)$$

where $V_s$ and $V_d$ indicate the volume of the discs in the swollen and dry state, respectively.

## *2.9. PS release tests*

Samples of PHOVA, P3VA35, 4VBC* and I3VA35 were individually incubated in 3 mL of 1X PBS (pH 7.4, 72 h) at either room temperature (25 ℃) or 37 ℃. The absorbance of the supernatant was measured at selected time points (2h, 24h, 48h, 72h, 96h) using a microplate reader (Varioskan LUX, Thermo Scientific). The measured values were subtracted from the absorbance value of the sample-free PBS solution and used to quantify the leaching of TB from the cured samples by building a calibration curve for TB. The calibration curve was generated in PBS by recording the absorbance of nine solutions of PBS supplemented with increasing concentration of TB (Figure S1, Supp. Inf.). The 96-hour mediums were freeze dried and resuspended in PBS for Liquid Chromatography-Mass Spectrometry (Supp. Inf.).

*2.10.ROS fluorometric assay*

2 mM of stock solution of 2',7'–dichlorodihydrofluorescein diacetate (DCFH-DA, Invitrogen) was prepared by dissolving 0.974 mg in 1 mL of absolute ethanol and was covered with aluminium foil to avoid contact with light, in line with previous reports [32]. 250 µL of the DCFH-DA solution were diluted with 250 µL of absolute ethanol and reacted with 2 mL NaOH 10mM in distilled water. The mixture was left at room temperature in the dark for 30 minutes and then neutralised with the addition of 10 mL PBS. The DCFH working solution was kept refrigerated and used for up to 5 days after preparation [33]. Either TB-supplemented solutions (600 µL in PBS), dry solid films (~ 4-5 mg) or PBS (negative control) were transferred to a 12-well plate. 600 µL of the DCFH solution was subsequently added to each well prior to sample incubation (dark controls) or irradiation with work light (~15 mW cm$^{-2}$). 100 µL aliquots were transferred to a black 96-well plate at selected irradiation time points (0, 2, 5, 8,10 and 20 minutes) and at 10 and 20 minutes for the dark controls. Subsequently, the intensity of the ROS-associated fluorescent signal was quantified using a Varioskan LUX Multimode microplate reader with 485 nm and 520 nm wavelengths for excitation and emission, respectively.

*2.11.Radical Scavenging Activity Assay*

The radical scavenging activity of UV-cured samples was determined via 2,2-diphenyl-1-picrylhydrazyl (DPPH) assay, following a previously described protocol [34]. A solution of DPPH (16.5 mg in 300 mL ethanol) was prepared and dry solid films (~ 4-5 mg) were incubated for 30 min in the dark. Afterwards, the absorbance of four different solutions for each sample was measured at 517 nm and the radical scavenging activity (%RScA) was calculated using the following equation:

$$\%RScA = \frac{(A-B)-(C-D)}{(A-B)} \quad (4)$$

where A is a mixture of 1.0 mL of DPPH solution and 1.0 mL of PBS, B is a mixture of 1.0 mL of ethanol and 1.0 mL of PBS, C is a mixture of 1.0 mL of DPPH solution and 1.0 mL of PBS containing the dry test sample, and D is a mixture of 1.0 mL of ethanol and 1.0 mL of PBS containing the dry test sample. The results are presented in Table S7 of the Supporting Information.

*2.12.Antibacterial tests in vitro*

*Bacterial culture.* Bacterial cultures were prepared overnight via aseptic addition of 5 mL of Mueller Hinton Broth (MH) media to a sterile universal tube. A sterilised culture loop was used to collect a single bacterial colony from an agar plate containing either *S. aureus* NCTC 8532 or *P. aeruginosa* NCTC 10332 bacterial strains. Each bacterial colony was added to the test tube, prior to overnight aerobic incubation at 37 ℃. An absorbance reading of 0.08 was considered to equate to 1.5 ×10$^{8}$ CFU mL$^{-1}$ according to the McFarland 0.5 standard used to calculate the number of bacterial cells in the overnight culture solution.

*Sample preparation*. UV-cured samples of PHOVA, P2VA39 and P2VA88 (Ø ~10 mm, $m_d$ ~ 15 mg) were disinfected for 15 min on each side via UV light exposure. Samples (n =3) were subsequently hydrated (30 min) in a 24-well plate containing 1 mL of sterile PBS in each well, prior to transfer to 1 mL of 1 ×10$^{6}$ CFU·mL$^{-1}$ suspension of either *S. aureus* NCTC 8532 or *P. aeruginosa* NCTC 10332.

*Light activation.* Well plates containing the samples were irradiated with a work light (~15 mW cm$^{-2}$) for 10 min and 20 min. Well plates were subsequently incubated aerobically for four hours at 37 ℃. Remaining live bacteria (CFU mL$^{-1}$) were calculated through the dilution plating method and overnight incubation.

*2.13.Cytotoxicity study in vitro*

*Cell culture*. L929 mouse fibroblasts (passage number 15) were cultured (37 ℃, 5 % $CO_2$) in a T-150 cell culture flask with alpha-MEM supplemented with 10 % FBS, 1 % P-S, and 1 % Gln.

*Contact cytotoxicity tests.* UV-cured dry discs of P2V39 were incubated in 70 vol% ethanol and subsequently air-dried in the dark prior to use. Ethanol-treated samples were individually transferred to the wells of a 96-well plate, prior to L929 cell seeding (1×10$^{4}$ cells per well). The samples were subsequently exposed to 10-min work light (~15 mW cm$^{-2}$), as described above.

*Cellular metabolic activity*. Following 24- and 72-hour cell culture, media were removed from each well and replaced with fresh media supplemented with 10 % of AlamarBlue reagent. The cells were then incubated for a further 2 h at 37 ℃. Fluorescence was measured using a Thermo Scientific Varioskan Lux plate reader with excitation at 560 nm and emission at 590 nm. Fluorescence readings were analysed using OriginPro 2023b software. Each sample was tested in triplicate. Cellular tolerability was assessed based on a viability threshold of 70 % (ISO 10993–5:2019).

### *2.14. Statistical analysis*

Data normality tests and statistical analysis were carried out using Origin software (version 2024b Academic). Statistical significance of normally distributed data sets was determined with one-way ANOVA corrected with Bonferroni test (∗: $p < 0.05$). Non-normally distributed datasets were analysed using Kruskal–Wallis followed by Dunn's pairwise comparisons with Benjamini–Hochberg correction (#: $p < 0.05$). Normally distributed data were presented as mean ±standard deviation. Non-normally distributed data were expressed as median values with error bars representing the distance from the median to first and third quartiles.

## 3. Results and Discussion

### *3.1. Characterisation of functionalised PVA products*

The synthesis of the functionalised PVA products and corresponding UV-cured co-networks is presented in Figure 1. The first step of the synthesis involves the esterification of the hydroxyl groups of PVA with tosyl chloride, establishing sulfonyl ester residues. The creation of this ester transforms the hydroxyl group to a good leaving group, facilitating a nucleophilic substitution reaction with the amine group of toluidine blue, and the grafting of PVA with the chromophore [35]. To enable co-curing functionality, the remaining secondary alcohol groups of the resulting product are subsequently esterified with the methacrylate group, through the reaction with MA. This secondary functionalisation is performed after the addition of TB groups given the high reactivity of MA, and to minimise potential steric hinderance limiting the introduction of the bulky phenothiazine residues. All the reactions take place in water with the addition of TEA, whereby the use of water as reaction medium was mandated by the future use of the polymer in biological environments. Although esterification reactions in aqueous alkaline environment are not favoured, the selected synthetic pathway results in the successful polymer functionalisation.

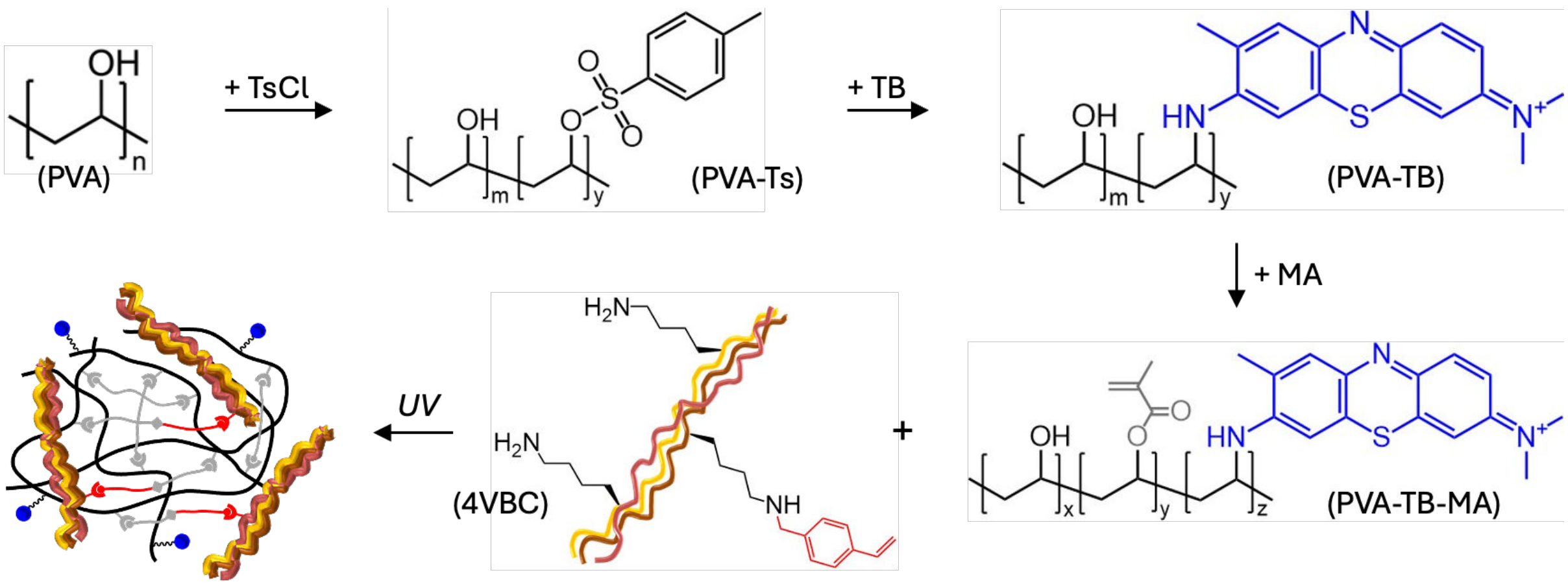

**Figure 1.** Design of a UV-cured co-network of a photodynamic and photocurable PVA derivative and 4-vinylbenzyl-functionalised type I collagen (4VBC). Tosylation of PVA (PVA-Ts) enables the grafting of toluidine blue (TB, blue; PVA-TB), followed by methacrylation of residual hydroxyl groups (grey) to yield TB-conjugated PVA methacrylate (PVA-TB-MA). The functionalised PVA is dissolved in a photoinitiator-supplemented solution containing 4VBC and exposed to UV light to induce free-radical crosslinking. Network formation proceeds predominantly through methacrylate-rich crosslinking and mixed methacrylate–4-vinylbenzyl crosslinking segments, resulting in a covalently integrated co-network of PVA and type I collagen. The formation of purely 4-vinylbenzyl-based crosslinking segments is expected to be kinetically disfavoured due to the lower radical propagation efficiency, restricted mobility, and lower molar content, of collagen-bound 4-vinylbenzyl groups compared with methacrylated PVA residues. (—): PVA chain; (): collagen molecule; (●): grafted TB; (●—◆—●): methacrylate–4-vinylbenzyl crosslinking segment; (●—●): methacrylated crosslinking segment; (): benzylic junction netpoint on type I collagen; (): ester-anchored methacrylate junction netpoint on PVA.

The chemical structure of the products and the degree of functionalisation for each substitution were verified through various spectroscopic techniques. Figure 2A shows the FTIR spectra of the native raw material and reacted products and shows the changes occurring to the chemical composition of the material after every step of the synthetic pathway. Characteristic FTIR signals of PVA include a crystallisation band at 1140 $cm^{-1}$, alongside broader signals at ca. 3200 $cm^{-1}$ and 2900 $cm^{-1}$ associated with the stretching vibrations of the O–H and C–H bonds, respectively [32].

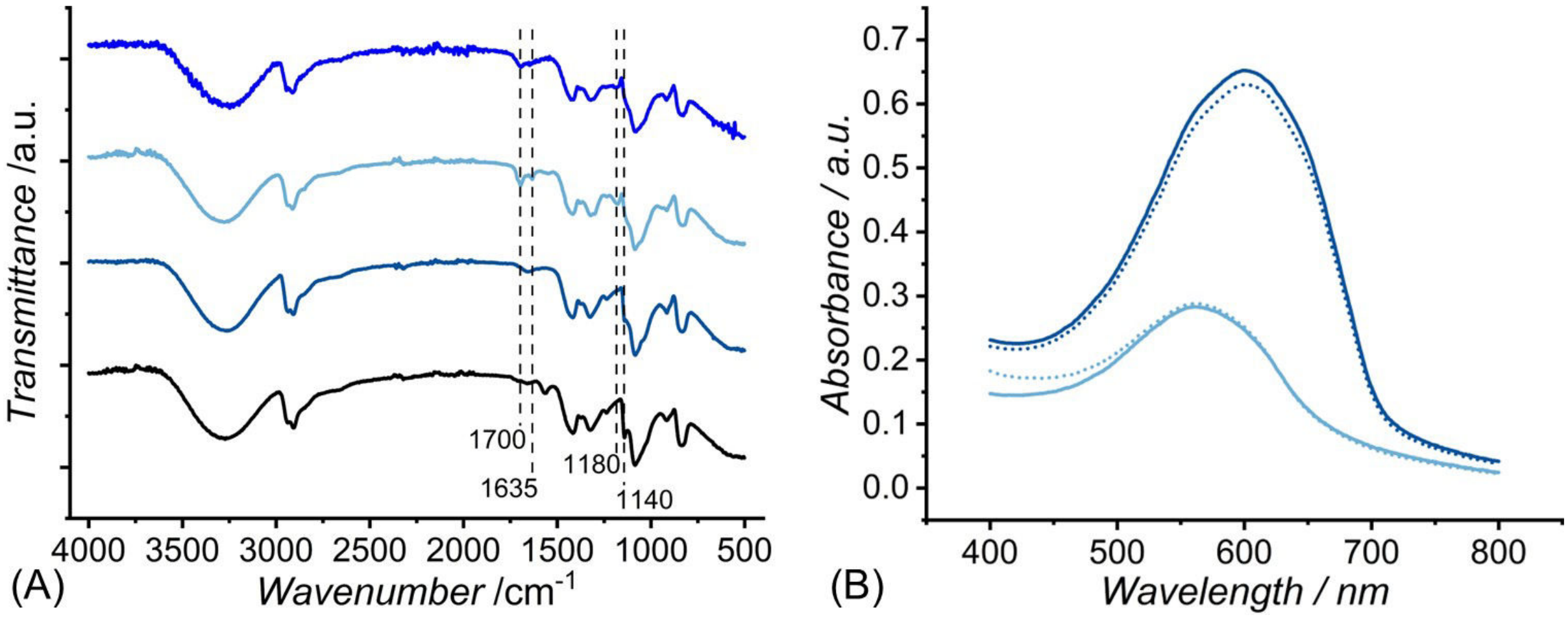

**Figure 2.** FTIR spectra (A) of dry samples of native and functionalised PVA, and UV-Vis spectra (B) of the corresponding solutions (10 mg·$mL^{-1}$, DMSO). (—): PVA, (—): PVA-TB, (—): PVA-TB-MA, (—): PHOVA. Two UV-Vis spectra (solid and dotted lines) are shown for solutions of PVA-TB and PVA-TB-MA (B).

The decreased intensity of the signal at 1140 $cm^{-1}$ in the spectra of TB-, MA- and UV-reacted products compared to the native polymer, is consistent with reduced PVA crystallinity following the introduction of phenothiazine residues, methacrylate residues, and UV-induced crosslinks.

The reaction of PVA with TB is associated with an additional band at ca. 1635 $cm^{-1}$, which is attributed to the vibrations of the C=C aromatic bonds of TB, in line with the covalent coupling of phenothiazine residues to the polymer backbone. Additional bands at 1180 and 1700 $cm^{-1}$ are detected in the FTIR spectrum of PVA-TB-MA, supporting the formation of the ester bond associated with the methacrylation of hydroxyl groups of PVA, while the 1635 $cm^{-1}$ absorption provides direct evidence of the introduction of the alkene group of MA. In the cured compound PHOVA, the absorption for the alkene bond is suppressed, indicating that covalent network formation is successfully achieved via exposure of the photoinitiator-supplemented solution of PVA-TB-MA to UV light.

TB incorporation was further investigated using UV-Vis spectroscopy following dissolution of PVA-TB and PVA-TB-MA in DMSO (Figure 2B). A blue shift in the absorption maxima was detected in the spectrum of PVA-TB-MA (560 nm) relative to PVA-TB (595 nm), and in both spectra relative to free TB ($\lambda_{max}$: 630-640 nm), providing indirect evidence of polymer methacrylation and TB grafting. Quantification of TB content was subsequently achieved by building a calibration curve of TB-supplemented solutions (Figure S1, Supp. Inf.), yielding values of 8.5 mg $g^{-1}$ and 3.5 mg $g^{-1}$ for PVA-TB and PVA-TB-MA, respectively (Table S2, Supp. Inf.). The lower TB content measured in PVA-TB-MA compared to PVA-TB is consistent with the removal of loosely associated (non-covalently bound) dye during the methacrylation step and subsequent purification.

Further to the investigations via FTIR and UV-Vis, the chemical structure of each functionalised PVA precursor and the final PVA product was explored by $^1$H-NMR spectroscopy. The characteristic signals of PVA at 3.84 ppm and 1.38 ppm, related to the methine and methylene groups, respectively, were detected in all spectra (Figure S3 A-C, Supporting Information). The additional aromatic signals, observed at ca. 7-8 ppm in the spectrum of PVA-Ts (Figure S3 B, Supporting Information), supported the grafting of tosylate groups to the polymer backbone. These signals were no longer observed in the spectrum of PVA-TB (Figure S3 C, Supporting Information), indicating consumption of tosylate residues during reaction with TB. Together with the UV-Vis data demonstrating retained TB content (Figure 2B), these observations are consistent with nucleophilic substitution and TB grafting. Although aromatic resonances were detected in the spectrum of native TB (Figure S3 D, Supporting Information), the corresponding phenothiazine signals could not be unambiguously resolved following reaction of PVA-Ts with TB (Figure S3 C, Supporting Information), in line with previous reports [36]. Other than the aromatic residues, the spectrum of PVA-TB indicated a low-intensity signal at 2.15 ppm, which was detected in a similar region in the spectrum of native TB, and potentially attributed to the presence of the methyl residue attached to the phenothiazine core. In comparison to PVA-TB, the spectrum of PVA-TB-MA revealed new signals at ca. 5-6 ppm, which were attributed to the alkene group of grafted MA residues (Figure 3). Integration of the respective peaks indicated a degree of MA grafting of 4 mol.%, in line with previous reports [32]. Interestingly, an additional signal was detected at 1.83 ppm, reflecting the presence of the methyl residues of MA. The low-intensity signal previously detected at 2.15 ppm in the spectrum of PVA-TB was still observed in the spectrum of PVA-TB-MA, suggesting retention of the species tentatively assigned to methyl groups associated with the phenothiazine structure. Taken together, the $^1$H-NMR spectra demonstrate successful tosylation of PVA, subsequent consumption of tosylate residues during reaction with TB, and final methacrylation of the resulting polymer. Direct quantification of TB incorporation by $^1$H-NMR was not feasible because phenothiazine resonances could not be unambiguously resolved in the TB-reacted polymers. These findings are consistent with the proposed functionalisation strategy (Figure 1) and complement the UV-Vis data used to confirm and quantify TB incorporation (Figure 2B).

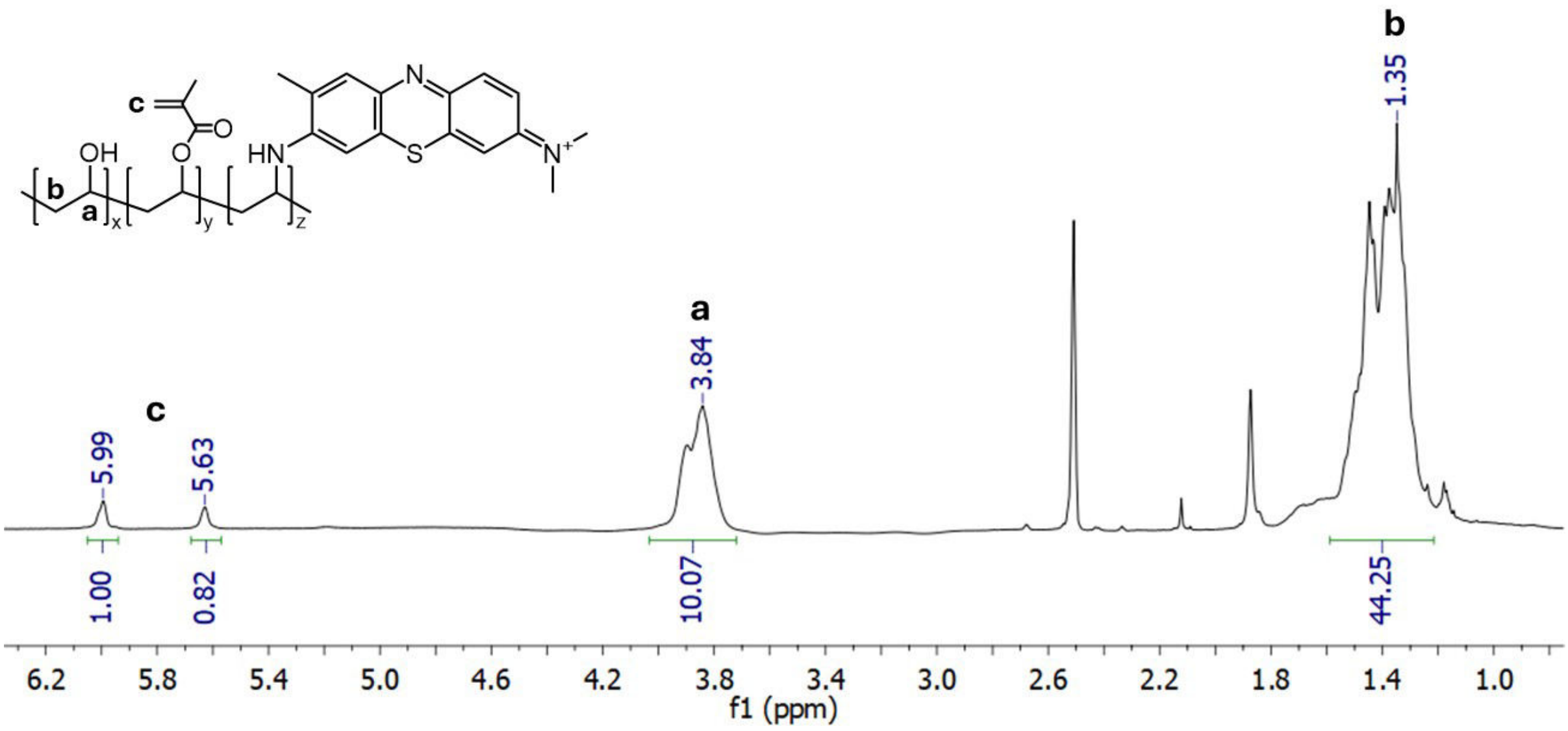

**Figure 3.** $^1$H-NMR spectrum (400 MHz) of PVA-TB-MA recorded in DMSO-$d_6$ (10 mg·mL$^{-1}$) at room temperature.

### *3.2. Chemical and thermal characterisation of UV-cured co-networks*

Having confirmed the functionalisation of PVA with both phenothiazine and methacrylate residues, the attention moved to the synthesis of the corresponding UV-cured homo-networks and PVA-collagen co-networks. A total of 12 co-cured networks were prepared by UV-curing a photo-initiator-loaded aqueous solution containing varying molar ratios of methacrylate residues of PVA and 4-vinylbenzyl residues of type I collagen (Table S3, Supp. Inf.). The resulting samples revealed a dark colour, comparable to the case of PHOVA and in contrast to the colour-less UV-cured 4-vinylbenzyl-functionalised rat tail collagen (Figure 4A-D). Samples were subjected to spectroscopic and thermal analyses aiming to explore their molecular configuration and establish structure-property relationships. Figure 4E depicts the FTIR spectra of a cured sample of PVA (PHOVA), a representative co-network (P3VA35) and UV-cured 4-vinylbenzyl-functionalised type I collagen (4VBC*). The co-network spectrum exhibits characteristic bands of both components, including the absorbance at ~1180 cm$^{-1}$ associated with methacrylate substitution, and a band around 2860 cm$^{-1}$ corresponding to C–H stretching vibrations, consistent with the presence of aromatic 4-vinylbenzyl residues. On the other hand, the spectral region around 1635 cm$^{-1}$ describes both C=C stretching of vinyl groups and the amide band of collagen, preventing direct assessment of vinyl conversion in the co-networks, unlike in the case of PHOVA (Figure 2A), where extensive consumption of unsaturated residues was confirmed.

Given the limitations of FTIR analysis in unambiguously confirming network formation in the presence of overlapping bands, samples of PHOVA, P3VA35 and 4VBC* were incubated in PBS for 96 hours at either 25 or 37 °C to assess the integrity of the UV-cured samples.

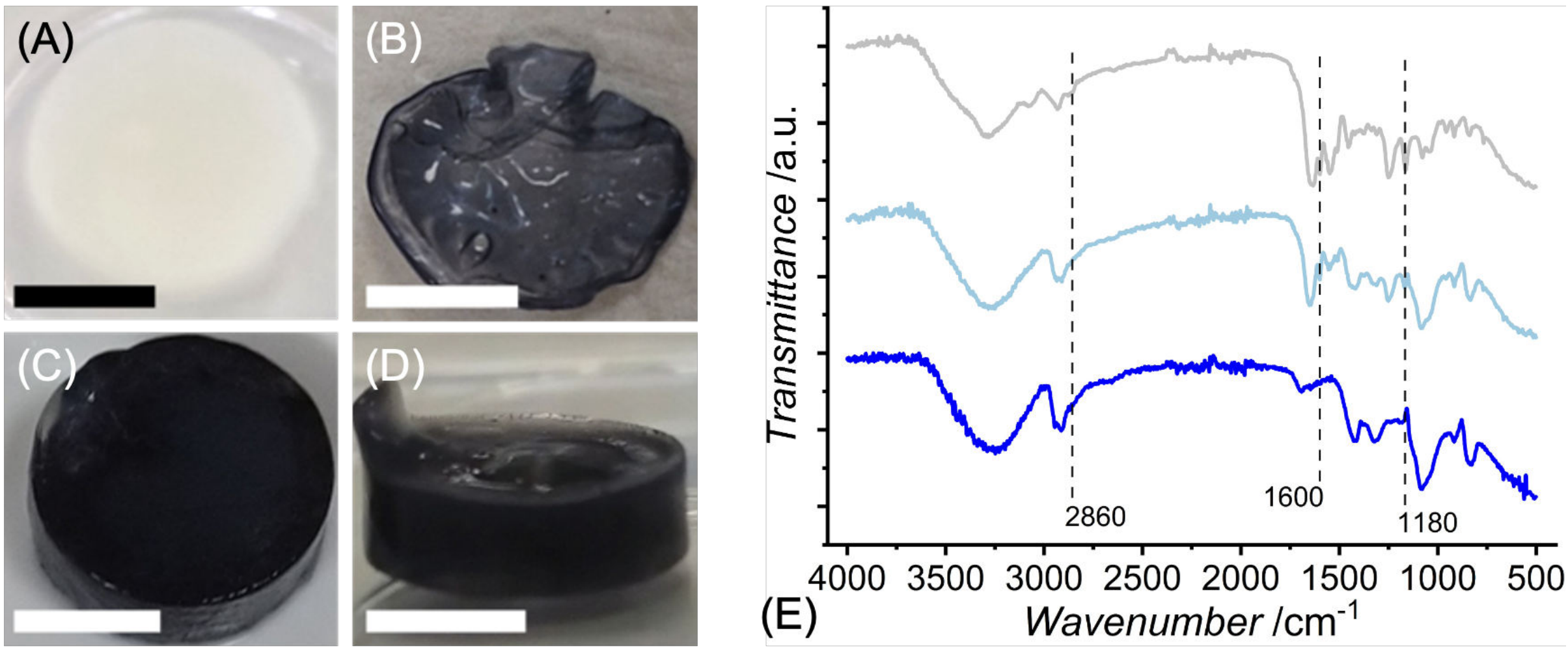

**Figure 4.** Appearance (A-D) and chemical characterisation (E) of UV-cured (co-)networks. (A-D): digital macrographs of UV-cured samples made of 4-vinybenzyl-functionalised rat tail collagen (4VBC*, A), TB-functionalised PVA methacrylate (PHOVA, B), and co-networks P2VA35 (C) and P3VA35 (D) made with an increasing concentration of the overall polymer blend. Images were captured after UV-curing of either the photoactive solution (A, C-D) or of the dry film (B). Scale bar ≈ 1 cm. (E): FTIR spectra of dry samples of PHOVA (—), P3VA35 (—) and 4VBC* (—).

Following incubation, recovered samples exhibited an average mass retention higher than 89 wt.% for PHOVA and 98 wt.% for the co-cured networks (Table S4, Supp. Inf.), consistent with the synthesis of a crosslinked co-network and in contrast to the complete dissolution of non-cured controls. UV-Vis spectroscopy of the supernatant revealed low absorbance values (< 0.005, $\lambda$ = 579 nm), equating to an average TB retention of more than 93 wt.% and a TB release of ca. 1 µg (96 h, 37 °C; Table S4, Supp. Inf.). Consistent with these findings, the supernatants remained clear and free of any visually detectable TB-induced staining in the medium (Figure S4, Supp. Inf.), supporting the successful conjugation of the phenothiazine residues to PVA. The low absorbance values were corroborated by LC-MS analysis of the supernatant collected at the final incubation time point (Figure S5, Supp. Inf.). Both mass and UV-Vis spectra (Figures S5 and S6, Supp. Inf.) also revealed signals associated with the release of residual photoinitiator species from the samples into the supernatant. These signals were substantially reduced following extended washing of the cured samples, indicating the efficacy of selected post-synthesis work-up in ensuring product purification (Table S1, Supp. Inf.).

Further to the incubation tests in wet environments, DSC analysis was performed on dehydrated samples in order to investigate the thermal properties and molecular organisation of the dry-state materials. This characterisation is relevant because the resulting materials are intended to be applied to the wound in the dry state prior to hydration by wound exudate. Consequently, DSC was used to examine thermal transitions associated with the dry precursor polymers and UV-cured materials rather than to describe the behaviour of the fully swollen hydrogels under physiological conditions. First heating scans were analysed for collagen-based samples (4VBC and 4VBC*) in order to capture thermal events associated with the native material architecture, including bound-water loss and collagen denaturation. Second heating scans were not considered for these materials because the initial thermal cycle irreversibly alters these features. In contrast, second heating scans were analysed for PVA-based materials to minimise the influence of thermal history and residual moisture. DSC thermograms revealed PVA-related glass transition and melting transitions in the dry cured co-networks (Figure 5A-E). Endothermic transitions associated with the collagen phase were also observed, most likely attributed to bound water loss ($T_{endo}$: 73–92 °C) and triple helix denaturation ($T_{endo}$: 174–182 °C) [37], both of which were detected in the DSC thermograms of the collagen homo-networks and respective functionalised precursor (Figure 5I-J). The glass transition temperature ($T_g$) revealed by the co-networks was lower than that of PVA-TB and PVA-TB-MA, suggesting plasticisation of the dry-state material due to residual bound water and the incorporation of the collagen phase. A melting transition at around 180 °C was observed for PHOVA (Figure 5F), while higher-temperature endothermic transitions were recorded for the co-networks. These observations suggest altered phase organisation and restricted chain mobility in the dry-state materials following incorporation of collagen. Other than the UV-cured samples, the DSC thermograms of PVA-TB and PVA-TB-MA (Figure 5G-H) revealed a $T_g$ at around 85 °C, consistent with values typically reported for PVA [38]. A melting transition was observed for both functionalised products, with a general shift towards lower temperatures compared to the native polymer ($T_m$: 220 °C), suggesting that the grafting of phenothiazine and methacrylate residues disrupt chain packing and reduces the crystallisation ability of PVA. Notably, PVA-TB-MA exhibited a lower $T_m$ than PVA-TB, consistent with the higher degree of functionalisation associated with methacrylate grafting relative to TB incorporation.

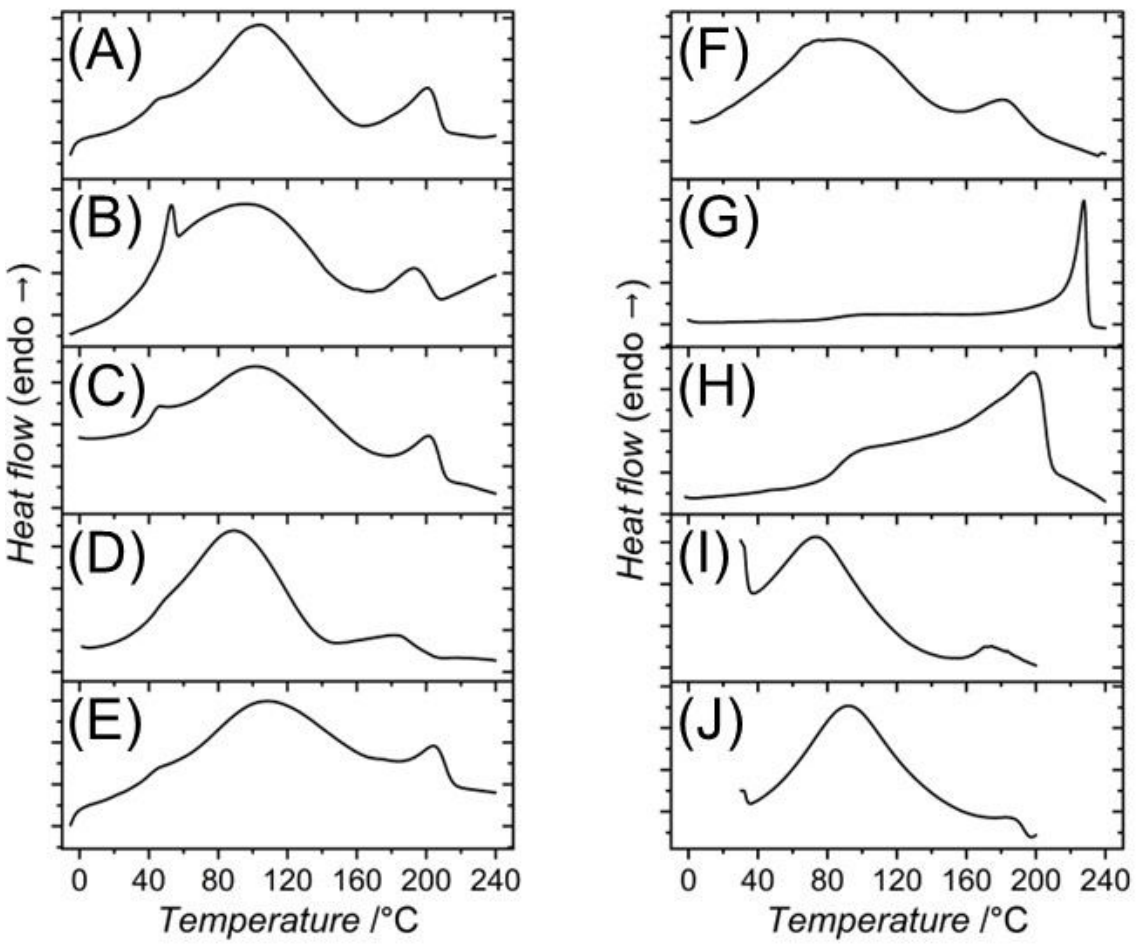

| Plot | Sample ID | $T_g$ /°C | $T_{endo}$ /°C | $T_m$ /°C |
|---|---|---|---|---|
| A | P2VA39 | 46 | 104 | 201 |
| B | P2VA53 | 53 | 96 | 193 |
| C | P2VA58 | 45 | 102 | 201 |
| D | P2VA88 | n.o. | 89 | 201 |
| E | P3VA35 | 45 | 108 | 204 |
| F | PHOVA | n.o. | 87 | 181 |
| G | PVA-TB | 85 | n.o. | 228 |
| H | PVA-TB-MA | 86 | n.o. | 199 |
| I | 4VBC* | n.o. | 73 (174) | n.o. |
| J | 4VBC | n.o. | 92 (182) | n.o. |

**Figure 5.** DSC thermograms (A-J, right) and corresponding transition temperatures (left) obtained from dry samples. First heating runs were analysed for collagen-based samples, whereas second heating runs were analysed for PVA-based samples. $T_g$: glass transition temperature; $T_{endo}$: endothermic transition temperature associated with either water loss or thermal decomposition (values in brackets); $T_m$: melting transition temperature. n.o.: not observed.

### *3.3. Physical characterisation of UV-cured co-networks*

The physical properties of the samples, namely swelling ratio (*SR*), volumetric swelling (*Q*) and gel content (*G*) were also assessed given their clinical relevance (Table 1). The significant swelling exhibited by the co-cured networks (*SR*: 498-935 wt.%; *Q*: 298-938 vol.%) can mainly be attributed to the introduction of 4-vinylbenzyl-functionalised collagen. Little weight increase and no volumetric expansion were observed with PHOVA, in line with other PVA-based crosslinked systems [39], and in contrast to the high swelling ratio exhibited by 4-vinylbenzyl-functionalised collagen homo-networks [31]. The average values of *G* were found to be higher than 89 wt.% across all UV-cured co-networks, while this did not exceed 80 wt% in PHOVA.

**Table 1.** Physical and mechanical properties of UV-cured samples. Swelling ratio (*SR*), volumetric swelling (*Q*) and gel content (*G*) were measured after 24-hour sample incubation in either distilled water at 37 °C (*SR* and *Q*) or 10 mM HCl at room temperature (*G*). Compression modulus ($E_c$), maximal compressive stress ($\sigma_{max}$) and ultimate compression ($\varepsilon_u$) were measured in the fully swollen state. Data are presented as average ± standard deviation (n=3-6). a-f: $p < 0.05$ (ANOVA, Bonferroni correction). n.o.: not observed; n.d.: not determined.

| Sample ID | *SR* /wt.% | *Q* /vol.% | *G* /wt.% | $E_c$ /kPa | $\sigma_{max}$ /kPa | $\varepsilon_u$ /% |
|---|---|---|---|---|---|---|
| P2VA28 | $503 \pm 9^{a,f}$ | $938 \pm 436^{a\text{-}d}$ | $89 \pm 3$ | $157 \pm 1$ | $109 \pm 17$ | $86 \pm 15$ |
| P2VA35 | $661 \pm 2^{b,f}$ | $310 \pm 93^{a}$ | $91 \pm 2$ | $247 \pm 100$ | $147 \pm 74$ | $72 \pm 20$ |
| P2VA39 | $625 \pm 169^{c,f}$ | $325 \pm 113^{b}$ | $99 \pm 13^{a}$ | $76 \pm 34^{a}$ | $51 \pm 35$ | $64 \pm 7$ |
| P2VA88 | $642 \pm 41^{d,f}$ | $606 \pm 171$ | $97 \pm 2$ | $132 \pm 82$ | $94 \pm 18$ | $64 \pm 13$ |
| P3VA35 | $498 \pm 28^{e,f}$ | $377 \pm 111^{c}$ | $97 \pm 1$ | $287 \pm 94^{a}$ | $149 \pm 39$ | $67 \pm 15$ |
| P3VA39 | $935 \pm 42^{a\text{-}f}$ | $298 \pm 34^{d}$ | $93 \pm 1$ | $148 \pm 73$ | $90 \pm 33$ | $74 \pm 4$ |
| PHOVA | $88 \pm 25^{f}$ | n.o. | $80 \pm 5^{a}$ | n.d. | n.d. | n.d. |

Given that the gel content reflects the fraction of insoluble matrix, these results support efficient network formation in the co-cured systems, likely enabled by the favourable cross-propagation between methacrylate-centred radicals and styrenic vinyl groups under UV-induced free radical conditions [40]. The increased *SR* observed in the co-networks, despite their high gel content, suggests that the incorporation of collagen promotes water uptake through increased hydrophilicity and reduced chain packing efficiency, in line with the initial hypothesis of this study. On the other hand, significantly decreased values of volumetric swelling were exhibited by samples obtained with increased methacrylate content (P2VA28: *Q*= 938±436 wt.% → P2VA88: *Q*= 606±171 vol.%; Table 1). These observations reflect an increased availability of reactive functional groups during curing, which is expected to enhance network connectivity and increase the effective crosslink density of the resulting networks.

To provide additional experimental evidence supporting the formation of a UV-cured co-network of PVA and type I collagen, interpenetrating network I3VA35 was prepared with the same composition as per sample P3VA35 (Table S3, Supp. Inf.) but using PVA-TB rather than PVA-TB-MA. This was aimed at generating an interpenetrating network, lacking methacrylate residues, on the PVA component and therefore designed to prevent covalent integration between the PVA and the collagen phases. A mean G of ca. 72 wt.% was exhibited by I3VA35 (Table S5, Supp. Inf.), while the mean values of mass retention and TB retention were both found to be below 80% (Table S6, Supp. Inf.) after 96-hour incubation *in vitro* (1X PBS, 25 °C). These data were lower than those measured with sample P3VA35 (Table 1), whereby a mean mass retention above 98% was shown under the same conditions (Table S4, Supp. Inf.). These results, together with the favourable cross-propagation between methacrylate-centred radicals and styrenic vinyl groups under UV-induced free radical conditions [41], provide additional experimental evidence consistent with covalent integration of the PVA and collagen phases within the UV-cured network. The substantially improved gel content, TB retention and mass retention observed for the MA-based formulations relative to the MA-free control are consistent with a more extensively crosslinked and structurally integrated co-network.

The mechanical properties of the water-equilibrated co-cured networks were subsequently measured (Table 1, Figure 6). Samples made from UV-curing solution containing an overall polymer concentration of 3 wt.% (P3VAXX) tended to exhibit higher compression modulus than their counterparts prepared with decreased overall polymer concentration and comparable ratios of MA to 4-vinylbenzyl residues (P2VAXX). This trend was further supported by tensile tests, whereby UV-cured co-networks P2VA39 and P3VA39 revealed a tensile modulus of 641 and 703 kPa, respectively, with strain at break values exceeding 80% in both samples (Figure S7, Supp. Inf.).

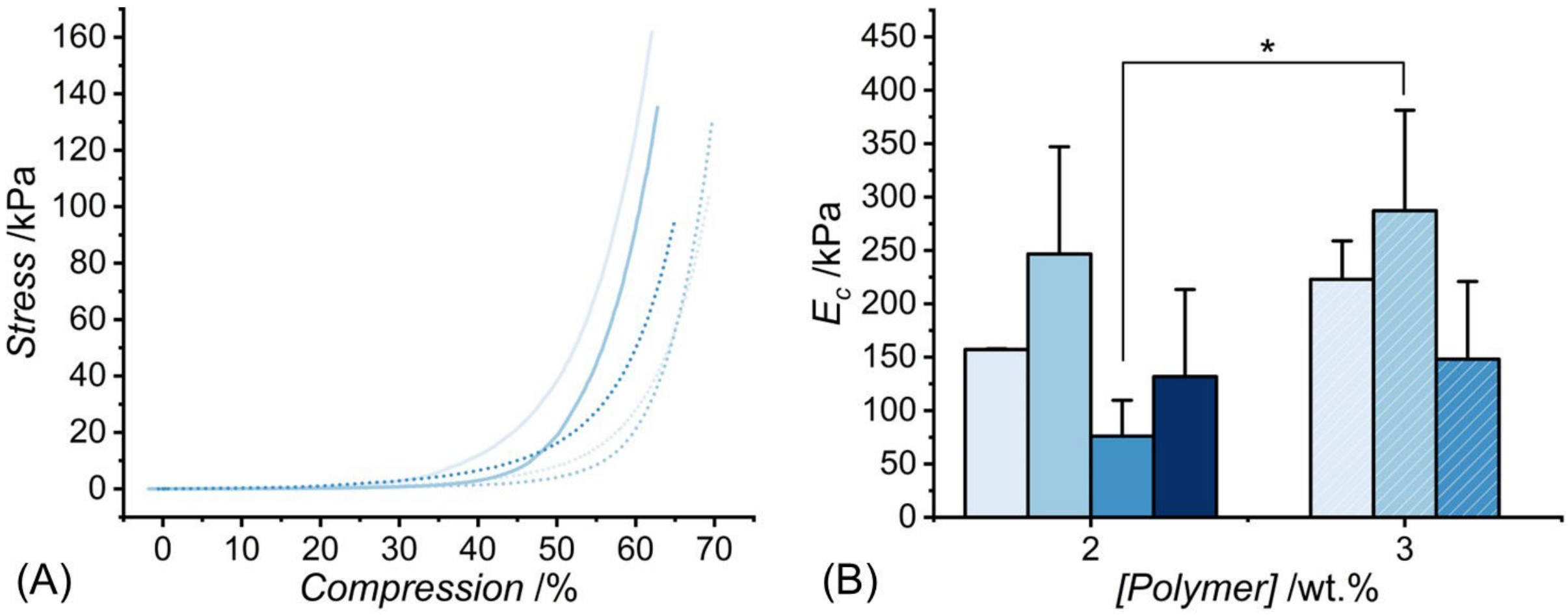

**Figure 6.** Compression measurements of UV-cured co-networks made of PVA-TB-MA and 4VBC-functionalised type I collagen. (A): Representative stress-compression curves recorded with P2VA28 (···), P3VA28 (—), P2VA35 (···), P3VA35 (—), P2VA39 (···). (B): compression modulus of samples made with varied concentration of the overall polymer. (□): P2VA28 (n=3); (□): P2VA35 (n=6); (□): P2VA39 (n=4); (■): P2VA88 (n=3); (□): P3VA28 (n=3); (□): P3VA35 (n=6); (□): P3VA39 (n=3). Data are presented as mean with positive standard deviation. Statistical significance was determined via one-way ANOVA with Bonferroni correction (∗: $p < 0.05$).

The elastic modulus of biomaterials is directly related to their resistance to elastic deformation in response to applied mechanical stress [41,42]. Consistent with the observed swelling behaviour, the higher compressive and tensile moduli exhibited by P3VAXX samples suggest that an increased overall polymer concentration in the UV-curing solution promotes the synthesis of mechanically strong networks. The biphasic, collagen-based composition of the UV-cured samples was also found to generate a nonlinear, *J*-shaped stress–strain response (Figure 6), typical of collagen-based materials. For this reason, the choice of the strain region used to quantify the elastic modulus is critical (corresponding to 10-15% of the maximal stress in this study) [26,31,37,43,44]. Moreover, the samples did not reach a failure point during compression testing, indicating high elasticity and the absence of compressive failure, unlike the case of UV-cured homo-networks of 4-vinylbenzyl-functionalised collagen.

### *3.4. Photodynamic activity*

Following characterisation of the chemical and physical properties, the photodynamic behaviour of the co-networks was investigated prior to antibacterial testing *in vitro*. The photodynamic response of the presented co-cured networks was evaluated following exposure to work light, with samples irradiated for 0-10 min and ROS production was assessed with the DCFH fluorometric assay (Figure 7 and Figure S8, Supp. Inf.) [45].

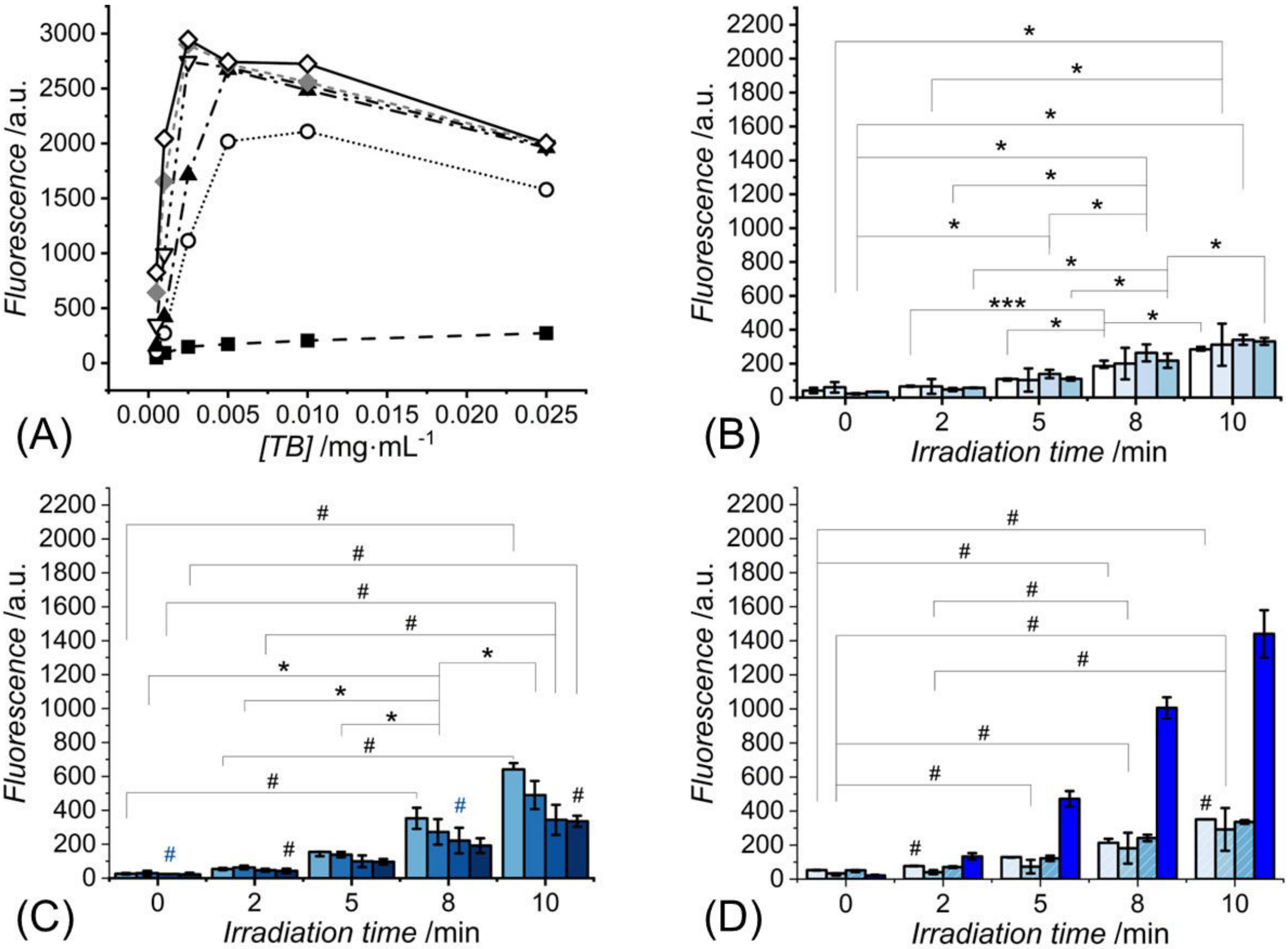

**Figure 7.** ROS generation determined via DCFH assay. (A): calibration curves obtained from aqueous solutions supplemented with increasing concentration of TB upon varying light exposure time. (– –■– –): 0 min; (··○··): 1 min; (–· ▲ –·): 2 min; (–· · ▽ –· ·): 5 min; (-- ◆ --): 8 min; (—◇—: 10 min). Data are shown as single points; lines are guidelines to the eye. (B-D): ROS-associated fluorescence generated by UV-cured samples made of either 4VBC, PHOVA or blends thereof, upon varying time duration of work light exposure. (□): 4VBC*; (□): P2VA28 (n=2-3); (□): P2VA32 (n=3); (□): P2VA35; (n=3); (□): P2VA39 (n=3); (□): P2VA44 (n=5); (■): P2VA53 (n=3-4); (■): P2VA88 (n=3); (□): P3VA28 (n=5); (□): P3VA35 (n=7); (□): P3VA39 (n=2); (■): PHOVA (n=2). Data are expressed as median values with error bars representing the distance from the median to first and third quartiles. Statistical significance of normally distributed datasets was assessed using one-way ANOVA with Bonferroni correction (*: $p < 0.05$; ***: $p < 0.001$). Non-normally distributed datasets were analysed using Kruskal–Wallis followed by Dunn's pairwise comparisons with Benjamini–Hochberg correction (#: $p < 0.05$).

Although previous studies have investigated irradiation times of up to 6 minutes [33], a wider range of irradiation time was tested in this study to establish structure-function relationships, given that the immobilisation of photosensitisers on hydrogels is known to generate ROS signals with increased stability [45]. Preliminary experiments showed an increasing ROS production with irradiation times of up to 10 minutes, followed by a plateau in signal intensity. Exposure times of up to 10 minutes were therefore investigated to test relatively short, clinically compliant sample illumination conditions and minimise cytotoxicity risks associated with high levels of ROS [46]. The trends in ROS generation observed for TB solutions, 4VBC* and PHOVA, as negative and positive controls, respectively, and the corresponding co-networks are reported in Figure 7. Consistent with the photodynamic response of TB solutions, the highest and lowest ROS signals were measured with PHOVA and 4VBC*, respectively, while intermediate signal intensities were recorded for the co-networks. The photodynamic activity of PHOVA and corresponding co-networks was further supported by DCFH assays conducted under dark conditions, whereby minimal ROS production was observed after both 10-min and 20-min incubation periods (Figure S8, Supp. Information).

Exposure time proved to significantly influence the photodynamic response of all samples, whereas the molar ratio between the photoactive constituents proved to have only a limited effect. The highest ROS signals were generated by PHOVA across all exposure times, while signal intensities measured for 4VBC* were of similar magnitude to those recorded with co-networks containing decreased PVA content (P2VA28-P2VA35), likely due to the photodynamic activity of the aromatic residues in 4VBC* [47]. Additionally, no significant differences in ROS production were observed for co-networks prepared with increased overall polymer concentration and constant molar ratios of the photoactive constituents. Since the TB content across these samples is relatively stable (Table S3, Supp. Inf.), the findings suggest that photosensitiser immobilisation plays a major role in ROS generation, whereby increased amounts of the photoactive precursors and reduction in free volume in the co-network may reduce TB accessibility and susceptibility to light activation. This behaviour may reflect the abiotic nature of the DCFH assay and the moderate radical scavenging activity exhibited by the UV-cured samples (Table S7, Supp. Inf.), which may limit the net accumulation of detectable ROS despite continued light exposure.

### *3.5. Antibacterial tests and cytotoxicity study in vitro*

Following ROS analysis, samples were subsequently evaluated *in vitro* against *S. aureus* and *P. aeruginosa*, as representative gram-positive and gram-negative chronic wound pathogens, respectively [47]. Co-networks showing the highest (P2VA39) and lowest (P2VA88) ROS activity were selected for further testing, alongside PHOVA and 4VBC* as the positive and negative controls, respectively. As presented in Figure 8A, both co-cured networks produced mean reduction of approximately 1.0-1.1 $\log_{10}$ in *S. aureus* viability following 10-min exposure to work light, while 1.9 $\log_{10}$ and <0.1 $\log_{10}$ mean reductions were measured with PHOVA and 4VBC* in the same experimental conditions, respectively.

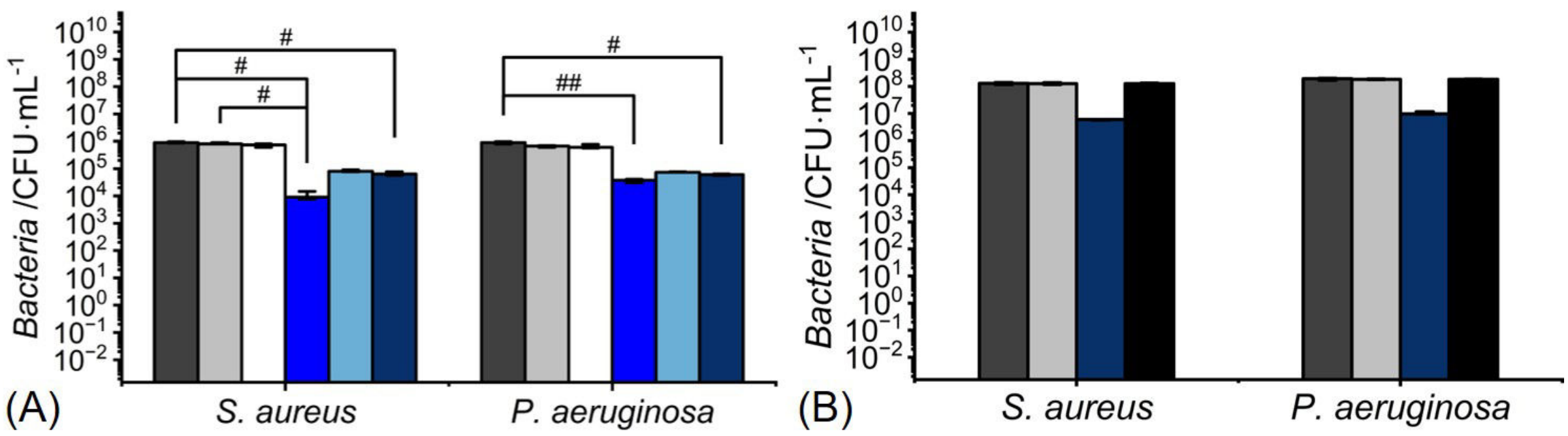

**Figure 8.** Antibacterial tests on UV-cured samples following work light exposure for either 10 min (A) or 20 min (B). Samples were cultured aerobically (4h, 37 °C) with either *S. aureus* or *P. aeruginosa*. (■): bacteria control; (□): light control; (□) 4VBC*; (■): PHOVA; (■): P2VA39; (■): P2VA88; (■): P2VA88 (dark control). Data are presented as median values with error bars representing the distance from the median to first and third quartiles (n = 3). Statistical significance was assessed using Kruskal–Wallis tests followed by Dunn's pairwise comparisons with Benjamini–Hochberg correction (#: $p < 0.05$, ##: $p < 0.01$).

Similar trends were observed when the same samples were tested with *P. aeruginosa*, indicating a significant antibacterial effect of these photodynamic co-networks against both gram-positive and gram-negative bacteria, despite the latter ones being notoriously harder to inactivate via classic PDT, due to the presence of an outer membrane.

Increasing the work light exposure time from 10 min to 20 min for sample P2VA88 resulted in median bacterial reductions of approximately 1.33 $\log_{10}$ and 1.28 $\log_{10}$ for *S. aureus* and *P. aeruginosa*, respectively, relative to the corresponding light-irradiated controls (Figure 8B). In contrast, no detectable antibacterial effect (0.00 $\log_{10}$ median reduction) was observed between dark controls of P2VA88 and either light-irradiated bacterial controls. These findings support that the antibacterial activity originated from a light-dependent photodynamic process, consistent with the ROS generation trends observed previously (Figure 7). The comparable reductions observed for *S. aureus* and *P. aeruginosa* are notable as they indicate a broad-spectrum antibacterial effect, contrasting typical PDT systems which predominantly affect Gram-positive bacteria [48]. Although antibacterial activity remained lower than that of PHOVA, these findings demonstrate light-dependent antibacterial efficacy and indicate that co-network performance may be further improved through optimisation of irradiation conditions.

Together with the antibacterial activity, it was important to assess the impact of sample illumination on mammalian cells, to confirm preservation of cell viability. Consequently, co-network P2VA39 (showing the highest ROS activity) was seeded with L929 murine fibroblasts prior to 10-min light exposure, followed by 24-hour culture and quantification of cellular metabolic activity. The Alamar Blue assay revealed a non-significant decrease in metabolic activity in cells cultured on the co-network with respect to those cultured on the tissue culture plastic control (Figure 9), corresponding to a relative signal intensity of ca. 90%, well above the 70% cytotoxicity threshold associated with quantitative *in vitro* tests [18]. These results therefore support the cellular tolerability of P2VA39, consistent with the behaviour of the corresponding homo-networks 4VBC* and PHOVA [9,32], and in contrast to the reported phototoxicity of free TB [49], indicating a beneficial effect of PS immobilisation within the co-network. Longer-term studies, including live/dead analysis and more comprehensive preclinical evaluation, will be required to fully assess the biocompatibility of these materials.

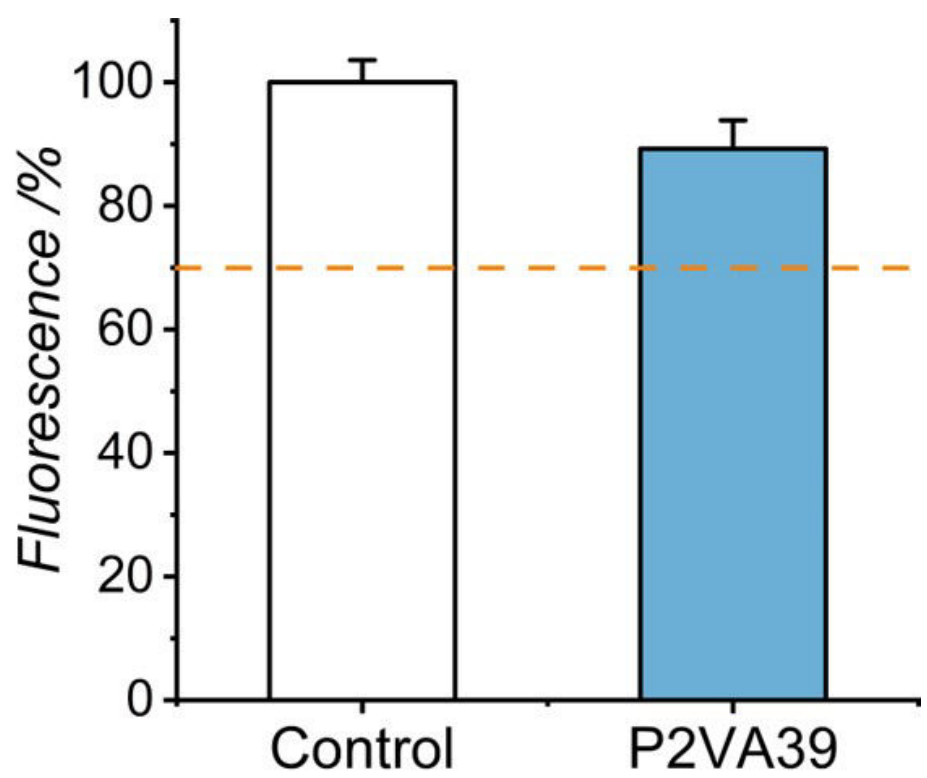

**Figure 9.** Relative metabolic activity of L929 fibroblasts measured via Alamar Blue assay following 24-hour cell culture onto UV-cured sample P2VA39 (blue column) or tissue culture plastic control (white column). Data are presented as mean with positive standard deviation (n=3) and normalised to the control (one-way ANOVA, $p > 0.05$). The dashed line indicates the threshold below which toxicity is expected.

## 4. Conclusions

UV-cured photodynamic co-networks were successfully developed from aqueous solutions containing 4-vinylbenzyl-functionalised rat tail collagen (4VBC), toluidine blue (TB)-conjugated polyvinyl alcohol methacrylate (PVA-TB-MA) and a water-soluble photoinitiator. The resulting co-networks displayed gel contents consistently above 89 wt.% as well as high mass retention (> 98 wt.%) and high TB retention (> 93 wt.%) following 96-hour incubation *in vitro* (pH 7.4, 37 ºC). These data were significantly higher than those obtained with an interpenetrating network control made of TB-conjugated PVA (lacking methacrylate residues) and 4VBC, indirectly supporting UV-induced covalent integration between methacrylate and styrenic functionalities within the co-network of PVA-TB-MA and 4VBC, as well as successful TB conjugation. The incorporation of 4VBC in the covalent co-network was key to achieving materials with significant mean swelling ratio (> 498 wt.%), driven by enhanced collagen-induced network hydrophilicity and reduced chain packing efficiency. Variation in polymer concentration and methacrylate content allowed tuning of volumetric swelling and wet-state mechanical properties, so that improved compressive moduli (up to ca. 280 kPa) and tensile moduli (> 640 kPa) were obtained while maintaining high extensibility (> 80%) and a characteristic nonlinear stress–compression behaviour typical of collagen-based systems. Photodynamic investigations demonstrated that ROS generation was strongly dependent on irradiation time, with statistically significant increases observed up to 10 min exposure. In contrast, variations in co-network composition and polymer concentration had a limited effect on ROS output, highlighting that immobilisation of TB within the polymer network influences its accessibility and photodynamic efficiency. The developed co-networks exhibited statistically significant antibacterial activity against both *S. aureus* and *P. aeruginosa*, inducing reductions in bacterial viability of at least one order of magnitude ($\geq$1 $\log_{10}$) upon 10-min work light exposure, while no bacterial reduction was observed in the corresponding dark conditions. The comparable response observed for Gram-positive and Gram-negative bacteria indicates broad-spectrum photodynamic activity, suggesting that the system may help overcome typical limitations associated with reduced susceptibility of Gram-negative strains. Importantly, cytotoxicity studies revealed that the co-network with the highest photodynamic activity maintained high mammalian cell metabolic activity (~90%), indicating cellular tolerance. This demonstrates that immobilisation of the photosensitiser effectively mitigates phototoxic effects commonly associated with freely diffusing TB, enabling a favourable balance between antibacterial efficacy and cytocompatibility. Overall, this work highlights the potential of collagen–PVA photodynamic co-networks as tunable, cell-tolerated platforms for light-activated antimicrobial applications with reduced risks of inducing bacterial resistance. The ability to modulate network architecture, mechanical properties, and photodynamic behaviour provides a valuable framework for the development of next-generation biomaterials with on-demand antibacterial activity for chronic wound healing applications.

**Acknowledgements**
The authors gratefully acknowledge the financial support from the Clothworkers' Company (London, UK). The technical assistance provided by Dr. Alex Heyam and Dr. Jeanine Williams is also gratefully acknowledged.

**Author contributions: CRediT**

N.M.: Data curation, Formal analysis, Investigation, Methodology, Visualization, Validation, Writing – original draft

C.B.: Formal analysis, Investigation, Methodology, Validation, Writing – review and editing

G.T.: Conceptualization, Formal analysis, Data curation, Funding acquisition, Project administration, Resources, Supervision, Writing – review and editing

## Graphical abstract

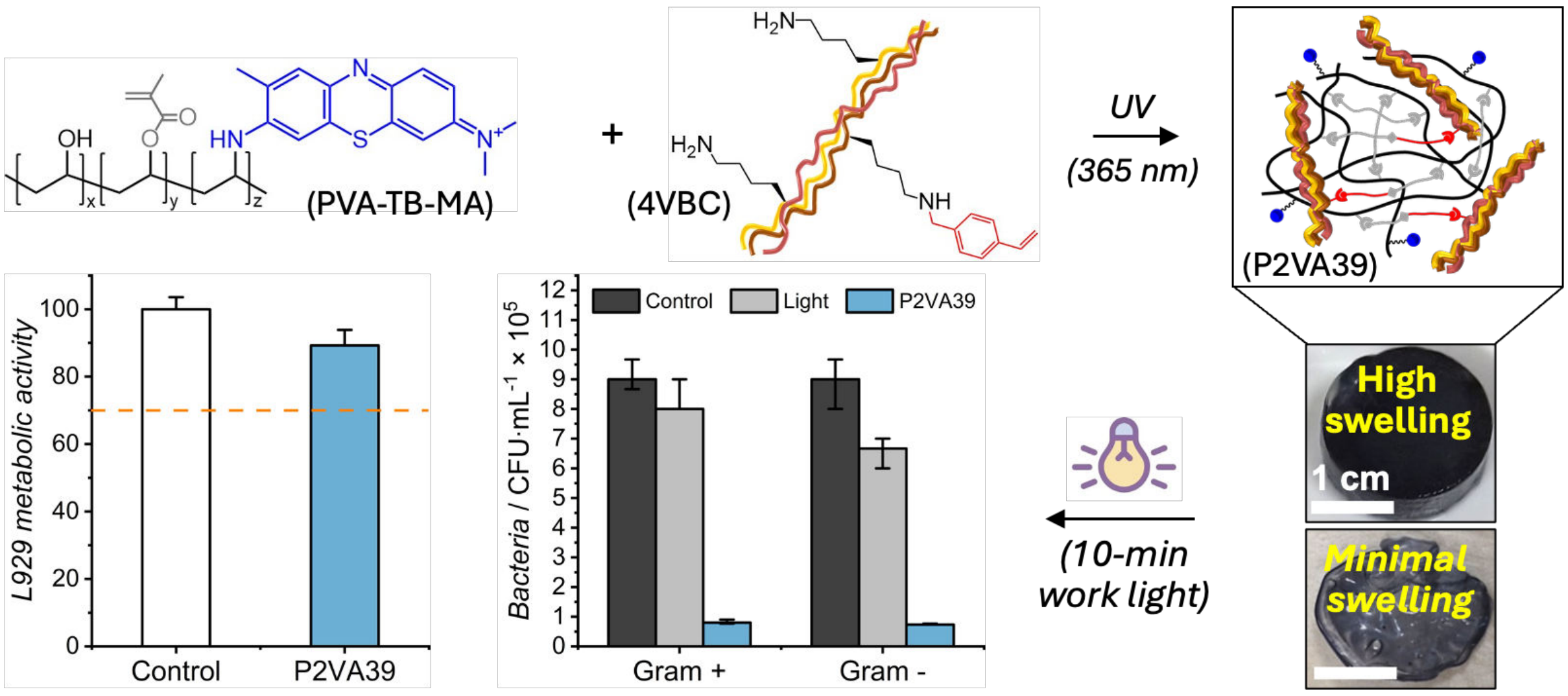

## References

[1] K.S. Priya, A. Gnanamani, N. Radhakrishnan, M. Babu, Healing potential of Datura alba on burn wounds in albino rats, *J. Ethnopharmacol.*, 83 (2002) 193–199. https://doi.org/10.1016/s0378-8741(02)00195-2

[2] S. Haldar, A. Sharma, S. Gupta, S. Chauhan, P. Roy, D. Lahiri, Bioengineered smart trilayer skin tissue substitute for efficient deep wound healing, *Mater. Sci. Eng. C*, 105, (2019) 110140. https://doi.org/10.1016/j.msec.2019.110140

[3] B. Silverberg, A structured approach to skin and soft tissue infections (SSTIs) in an ambulatory setting. *Clin. Pract.*, 11 (2021) 65–74. https://doi.org/10.3390/clinpract11010011

[4] P. Ganguly, E.A. Jones, P.V. Giannoudis, G. Tronci. Antibiotic-Free Strategies for Managing Antimicrobial-Resistant Infections. *Pharmaceutics*, 18 (2026) 1136. https://doi.org/10.3390/pharmaceutics18091136

[5] J. Pei, C.P. Palanisamy, P. Alugoju, N.V.A. Anthikapalli, P.M. Natarajan, V.R. Umapathy, B. Swamikannu, S. Jayaraman, P. Rajagopal, S. Poompradub, A Comprehensive Review on Bio-Based Materials for Chronic Diabetic Wounds, *Molecules*, 28 (2023) 604. https://doi.org/10.3390/molecules28020604

[6] V. Gounden, M. Singh, Hydrogels and Wound Healing: Current and Future Prospects, *Gels*, 10 (2024) 43. https://doi.org/10.3390/gels10010043

[7] W.-S. Liu, Y. Liu, J. Gao, H. Zheng, Z.M. Lu, M. Li, Biomembrane-Based Nanostructure- and Microstructure-Loaded Hydrogels for Promoting Chronic Wound Healing, *Int. J. Nanomed.*, 18 (2023) 385-411. https://doi.org/10.2147/ijn.s387382

[8] M. Ansari, A Darvishi, A review of the current state of natural biomaterials in wound healing applications, *Front. Bioeng. Biotechnol.*, 12 (2024) 1309541. https://doi.org/10.3389/fbioe.2024.1309541

[9] C. Brooker, G. Tronci, A collagen-based theranostic wound dressing with visual, long-lasting infection detection capability, *Int. J. Biol. Macromol.*, 236 (2023) 123866. https://doi.org/10.1016/j.ijbiomac.2023.123866

[10] S.M. Nagle, K.A. Stevens, S.C. Wilbraham, Wound assessment In StatPearls [Internet], StatPearls Publishing: Treasure Island, FL, USA, 2023

[11] E.A. Gantwerker, D.B.; Hom, Skin: Histology and physiology of wound healing, *Facial Plast. Surg. Clin. N. Am.*, 19 (2011) 441–453. https://doi.org/10.1016/j.fsc.2011.06.009

[12] J. Dissemond, P. Chadwick, D. Weir, P.; Alves, K. Isoherranen, J.L. Lázaro Martínez, T. Swanson, A. Gledhill, M. Malone, M.O.I.S.T. Concept for the Local Therapy of Chronic Wounds: An International Update, *Int. J. Low. Extrem. Wounds*, 25 (2026) 344-352. https://doi.org/10.1177/15347346241245159

[13] J.L. Daristotle, L.W. Lau, M. Erdi, J. Hunter, A. Djoum, P. Srinivasan, X. Wu, M. Basu, O.B. Ayyub, A.D. Sandler, P. Kofinas, 2020. Sprayable and biodegradable, intrinsically adhesive wound dressing with antimicrobial properties, *Bioeng. Transl. Med.*, 5, e10149. https://doi.org/10.1002/btm2.10149

[14] B. Sheokand, M. Vats, A. Kumar, C.M. Srivastava, I. Bahadur, S.R. Pathak, Natural polymers used in the dressing materials for wound healing: Past, present and future, *J. Polym. Sci*, 61 (2023) 1389–1414. https://doi.org/10.1002/pol.20220734

[15] J. Xiang, L. Shen, Y. Hong, Status and future scope of hydrogels in wound healing: Synthesis, materials and evaluation, *Eur. Polym. J.*, 130 (2020) 109609. https://doi.org/10.1016/j.eurpolymj.2020.109609

[16] F. Fan, S. Saha, D. Hanjaya-Putra, Biomimetic Hydrogels to Promote Wound Healing, *Front. Bioeng. Biotechnol.*, 9 (2021) 718377. https://doi.org/10.3389/fbioe.2021.718377

[17] G. Tronci, J. Yin, R.A. Holmes, H. Liang, S.J. Russell, D.J. Wood, Protease-sensitive atelocollagen hydrogels promote healing in a diabetic wound model, *J. Mater. Chem. B*, 4 (2016) 7249. https://doi.org/10.1039/c6tb02268e

[18] A. Contreras, M.J. Raxworthy, S. Wood, G. Tronci, Hydrolytic Degradability, Cell Tolerance and On-Demand Antibacterial Effect of Electrospun Photodynamically Active Fibres, *Pharmaceutics*, 12 (2020) 711. https://doi.org/10.3390/pharmaceutics12080711

[19] A.K. Surur, A.B. De Oliveira, S.R. De Annunzio, T.M. Ferrisse, C.R. Fontana, Bacterial resistance to antimicrobial photodynamic therapy: a critical update, *J. Photochem. Photobiol. B*, 255 (2024) 112905. https://doi.org/10.1016/j.jphotobiol.2024.112905

[20] R. Ding, X. Liu, X. Zhao, Q. Sun, Y. Cheng, A. Li, D. Pei, G. He, Membrane-anchoring selenophene viologens for antibacterial photodynamic therapy against periodontitis via restoring subgingival flora and alleviating inflammation, *Biomaterials*, 307 (2024) 122536. https://doi.org/10.1016/j.biomaterials.2024.122536

[21] H. Chen, S. Li, M. Wu Kenry, Z. Huang, C.S. Lee, B. Liu, Membrane-Anchoring Photosensitizer with Aggregation-Induced Emission Characteristics for Combating Multidrug-Resistant Bacteria, *Angew. Chem.*, 59

(2020) 632. https://doi.org/10.1002/anie.201907343
[22] D. Dipayan, P. Payel, Environmental impact of silver nanoparticles and its sustainable mitigation by novel approach of green chemistry, *Plant Nano Biol.*, 14 (2025) 100210. https://doi.org/10.1016/j.plana.2025.100210
[23] M.L. Chinta, P.K. Gandam, S.R.; Parcha, "Natural polymer based hydrogel systems for wound management" in Natural polymers in wound healing and repair, Elsevier, Netherlands, 2022, 129–165
[24] M. Muchová, L. Münster, Z. Capáková, V. Mikulcová, I. Kuřitka, J. Vícha, Design of dialdehyde cellulose cross-linked poly (vinyl alcohol) hydrogels for transdermal drug delivery and wound dressings, *Mater. Sci. Eng. C*, 116 (2020) 111242. https://doi.org/10.1016/j.msec.2020.111242
[25] S. Liu, D. Li, Y. Wang, G. Zhou, K. Ge, L. Jiang, Adhesive, antibacterial and double cross-linked carboxylated polyvinyl alcohol/chitosan hydrogel to enhance dynamic skin wound healing, *Int. J. Biol. Macromol.*, 228 (2023) 744–753. https://doi.org/10.1016/j.ijbiomac.2022.12.169
[26] G. Tronci, Advanced Textiles for Wound Care (Second Edition), Chapter 3, 363-389, 2019
[27] G. Tronci, J. Duch, D. Wood S. Russell, P. Castan, 2026. Hyfacol Skin Substitutes – Durability With Integrity, *Wound Repair Regen.*, 34, e70145. https://doi.org/10.1111/wrr.70145
[28] N. Raina, R. Rani, R. Pahwa, M. Gupta, Biopolymers and treatment strategies for wound healing: an insight view, *Int. J.Polym. Mater. Polym. Biomater.*, 71 (2022) 359–375. https://doi.org/10.3390/polym16091280
[29] G. Tronci, S.J. Russell, D.J.; Wood, Photo-active collagen systems with controlled triple helix architecture, *J. Mater. Chem. B*, 1 (2013) 3705–3715. https://doi.org/10.1039/c3tb20720j
[30] N. Rajan, J. Habermehl, M.F. Coté, C.J. Doillon, D. Mantovani, Preparation of ready-to-use, storable and reconstituted type I collagen from rat tail tendon for tissue engineering applications, *Nat. Protoc.*, 1 (2006) 2753-2758. https://doi.org/10.1038/nprot.2006.430
[31] C. Brooker, G. Tronci, Effect of Mammalian Tissue Source on the Molecular and Macroscopic Characteristics of UV-Cured Type I Collagen Hydrogel Networks, *Prosthesis*, 4 (2022) 1–14. https://doi.org/10.3390/prosthesis4010001
[32] M. Li, C. Brooker, R. Ambike, Z. Gao, P. Thornton, T. Do, G. Tronci, Photodynamic, UV-curable and fibre-forming polyvinyl alcohol derivative with broad processability and staining-free antibacterial capability, *Eur. Polym. J.*, 228 (2025) 113794. https://doi.org/10.1016/j.eurpolymj.2025.113794
[33] S. Jiang, B. Su, G. Li, X. Liu, P. Liu, M. Yang, J. Xiao, X. Huang, D. Wei, J. Sun, J. Ding, C. Wu, H. Fan, Emodin-derived red-emissive carbon dots: Light-driven ROS generation and antioxidant activities in non-illuminative regimes, *Biomaterials*, 324 (2026) 123496. https://doi.org/10.1016/j.biomaterials.2025.123496
[34] N. Guldbrandsen, M. de Mieri, M. Gupta, E. Liakou, H. Pratsinis, D. Kletsas, E. Chaita, N. Aligiannis, A.-L. Skaltsounis and M. Hamburger, Screening of Panamanian Plants for Cosmetic Properties, and HPLC-Based Identification of Constituents with Antioxidant and UV-B Protecting Activities, *Scientia Pharmaceutica*, 83 (2015) 177-190. https://doi.org/10.3797/scipharm.1409-12
[35] B. Wang, J. Wu, Y. Wu, C. Chen, F. Zou, A. Wang, H. Wu, Z. Hu, Z. Jiang, Q. Liu, W. Wang, Y. Zhang, F. Liu, M. Zhao, J. Hu, T. Huang, J. Ge, L. Wang, T. Ren, Y. Wang, J. Liu, Q. Liu, Discovery of 4-(((4-(5-chloro-2-(((1s,4s)-4-((2-methoxyethyl)amino)cyclohexyl)amino)pyridin-4-yl)thiazol-2-yl)amino)methyl)tetrahydro-2H-pyran-4-carbonitrile (JSH-150) as a novel highly selective and potent CDK9 kinase inhibitor, *Eur. J. Med. Chem.*, 158 (2018) 896-916. https://doi.org/10.1016/j.ejmech.2018.09.025
[36] J. Robinson-Duggon, N. Pizarro, G. Gunther, D. Zúñiga-Núñez, A.M. Edwards, A. Greer, D. Fuentealba, Fatty Acid Conjugates of Toluidine Blue O as Amphiphilic Photosensitizers: Synthesis, Solubility, Photophysics and Photochemical Properties, *Photochem. Photobiol.*, 97 (2021) 71-79. https://doi.org/10.1111/php.13304
[37] S.R. Sudhamani, M.S. Prasad, K. Udaya Sankar, DSC and FTIR studies on Gellan and Polyvinyl alcohol (PVA) blend films, *Food Hydrocoll.*, 17 (2003) 245-250. https://doi.org/10.1016/S0268-005X(02)00057-7
[38] J. Gorman, C. Brooker, X. Li, G. Tronci, An infection-responsive collagen-based wet-spun textile fibre for wound monitoring, *Polymer*, 339 (2025) 129113. https://doi.org/10.1016/j.polymer.2025.129113
[39] L. Lambertini, G. Coccarelli, E. Toto, M.G. Santonicola, S. Laurenzi, Poly(vinyl alcohol) gels cross-linked by boric acid for radiation protection of astronauts, *Acta Astronaut.*, 221 (2024) 142-154. https://doi.org/10.1016/j.actaastro.2024.05.020
[40] T.P. Davis, K.F. O'Driscoll, M.C. Piton, M.A. Winnik, Copolymerization propagation kinetics of styrene with alkyl methacrylates, *Macromolecules*, 23 (1990) 2113-2119. https://doi.org/10.1021/ma00210a003
[41] M. Kahl, D. Schneidereit, N. Bock, O. Friedrich, D.W. Hutmacher, C. Meinert, *MechAnalyze*: An Algorithm

for Standardization and Automation of Compression Test Analysis, *Tissue Eng Part C Methods*, 27 (2021) 529-542. https://doi.org/10.1089/ten.tec.2021.0170
[42] L. Vincent, A.J. Engler, Chapter 5.504 - Effect of Substrate Modulus on Cell Function and Differentiation in Comprehensive Biomaterials, Editor(s): Paul Ducheyne, Elsevier, 2011, 51-63
[43] H. Liang, J. Yin, K. Man, X.B. Yang, E. Calciolari, N. Donos, S.J. Russell, D.J. Wood, G. Tronci, A long-lasting guided bone regeneration membrane from sequentially functionalised photoactive atelocollagen, *Acta Biomater.*, 140 (2022) 190-205. https://doi.org/10.1016/j.actbio.2021.12.004
[44] R. Zhang, C. Brooker, L.L.E. Whitehouse, N.H. Thomson, D. Wood, G. Tronci, Mechanical and suture-holding properties of a UV-cured atelocollagen membrane with varied crosslinked architecture, *Biomed. Mater.*, 19 (2024) 065036. https://doi.org/10.1088/1748-605X/ad8828
[45] B. Mai, M. Jia, S. Liu, Z. Sheng, M. Li, Y. Gao, X. Wang, Q. Liu, P. Wang, Smart Hydrogel-Based DVDMS/bFGF Nanohybrids for Antibacterial Phototherapy with Multiple Damaging Sites and Accelerated Wound Healing, *ACS Appl. Mater. Interfaces*, 12 (2020) 10156-10169. https://doi.org/10.1021/acsami.0c00298
[46] K. Khorsandi, R. Hosseinzadeh, H. Esfahani, K. Zandsalimi, F. Khatibi Shahidi, H. Abrahamse, Accelerating skin regeneration and wound healing by controlled ROS from photodynamic treatment, *Inflamm. Regener.*, 42 (2022) 40. https://doi.org/10.1186/s41232-022-00226-6
[47] T.P. Kubrak, P. Kołodziej, J. Sawicki, A. Mazur, K. Koziorowska, D. Aebisher, Some Natural Photosensitizers and Their Medicinal Properties for Use in Photodynamic Therapy, *Molecules*, 17 (2022) 1192. https://doi.org/10.3390/molecules27041192
[48] M.R. Hamblin, T. Hasan, Photodynamic therapy: a new antimicrobial approach to infectious disease?, *Photochem. Photobiol. Sci.*, 3 (2004) 436-450. https://doi.org/10.1039/b311900a
[49] J.F. Tremblay, S. Dussault, G. Viau, F. Gad, M. Boushira, R. Bissonnette, Photodynamic therapy with toluidine blue in Jurkat cells: cytotoxicity, subcellular localization and apoptosis induction, *Photochem. Photobiol. Sci.*, 1 (2002) 852–856. https://doi.org/10.1039/b204385h

**Supporting Information**

# UV-cured PVA-collagen co-network with staining-free photodynamic activity against chronic wound bacteria

Nikitia Mexia,[1] Charles Brooker,[1,2] Giuseppe Tronci[1,2]

[1]Clothworkers Centre for Textile Materials Innovation for Healthcare (CCTMIH), School of Design, University of Leeds, Leeds, LS2 9JT, United Kingdom
[2]School of Dentistry, St. James's University Hospital, University of Leeds, LS9 7TF, United Kingdom

## Calibration curves of TB

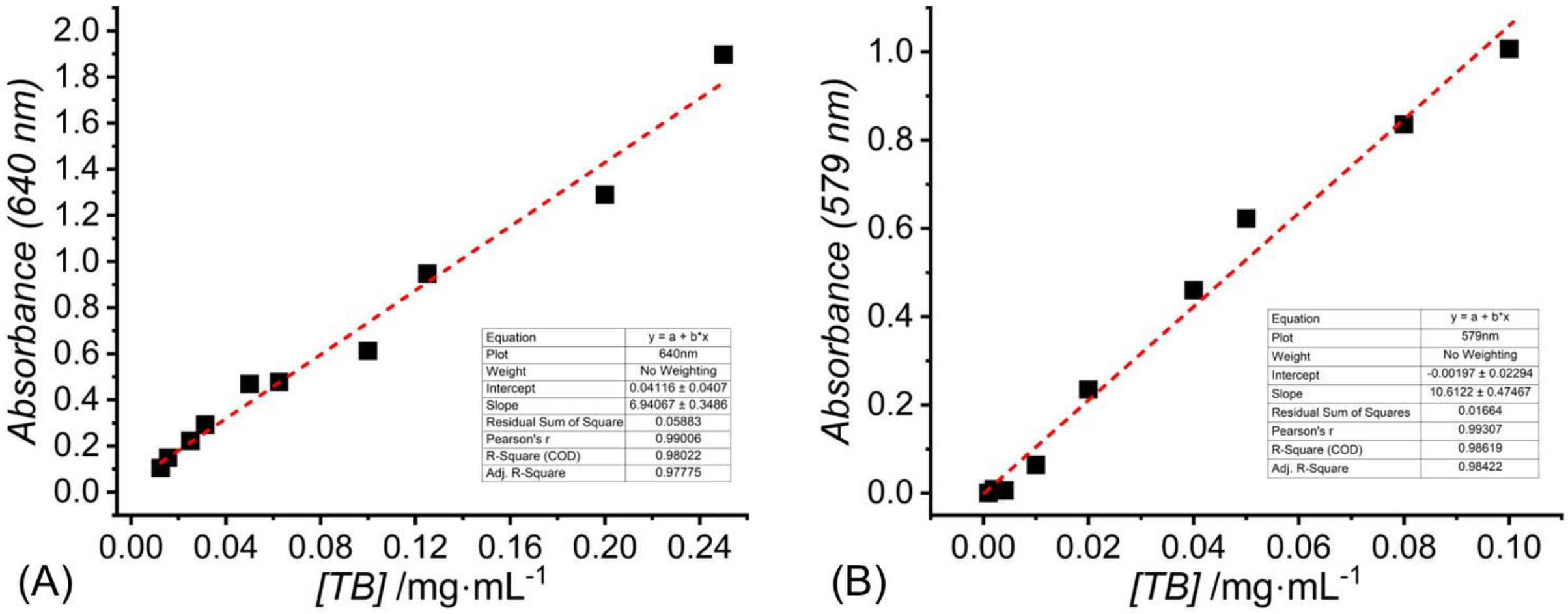

**Figure S1**. TB calibration curves recorded by UV-Vis spectroscopy at room temperature. TB was dissolved in either DMSO (A) or PBS (B) at increasing concentrations. Experimental absorbance data (■) were fitted by linear regression (---). The fitted equations and corresponding coefficients of determination ($R^2$) are reported within each plot.

## 2,4,6-trinitrobenzenesulfonic acid assay of collagen

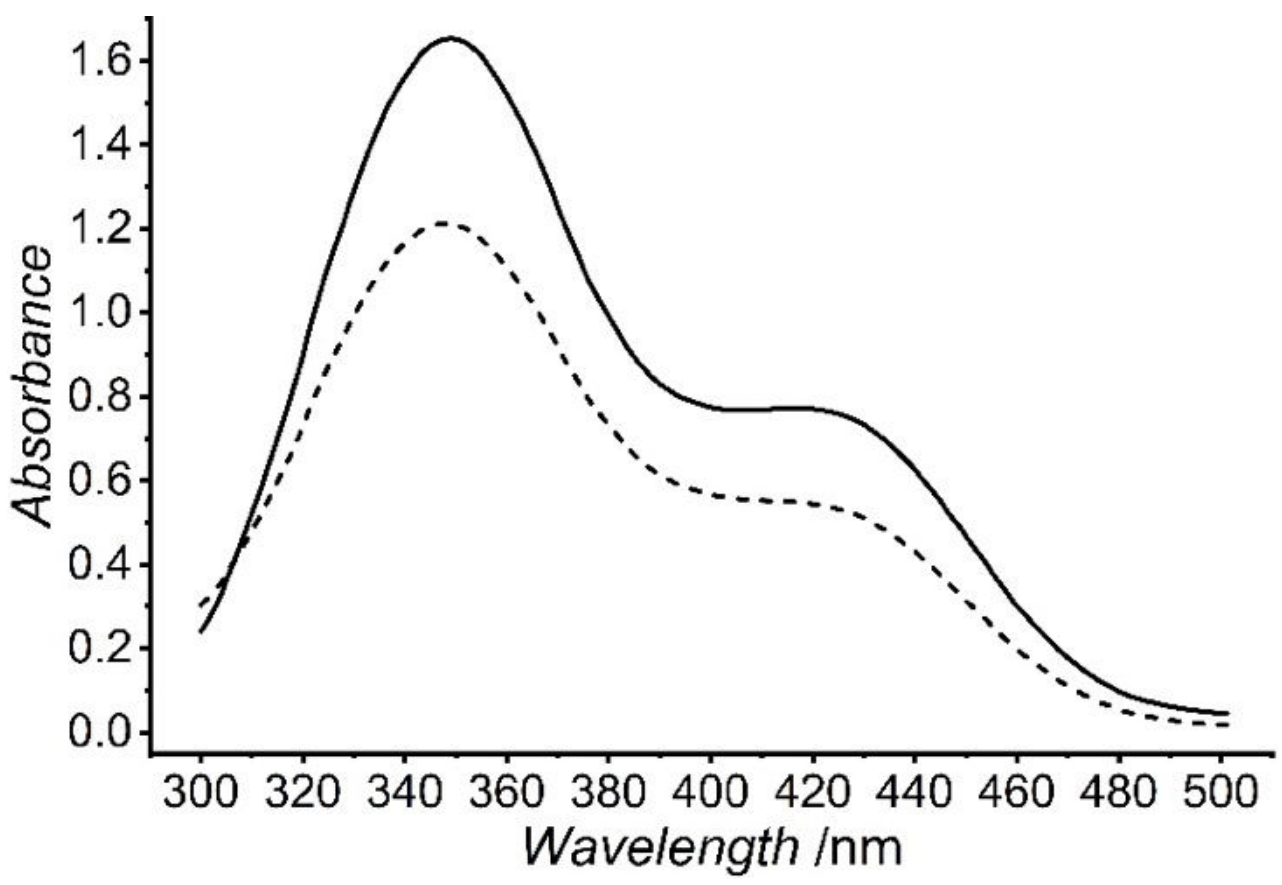

**Figure S2.** Representative UV-Vis spectra of native type I collagen (—) and 4-vinylbenzyl-functionalised collagen (---) following 2,4,6-trinitrobenzenesulfonic acid (TNBS) assay. The absorbance at 346 nm decreases following reaction with 4-vinylbenzyl chloride, indicating the consumption of collagen primary amino groups. Quantification of the corresponding absorbance values enabled determination of the degree of functionalisation ($F$ = 25 ± 2 mol.%, n=9).

## Washing procedure

**Table S1**. Post-synthesis purification procedure involving extended incubations in either acidic or neutral environments, as well as a final distilled water-ethanol dehydration series.

| Incubation step | Medium | Volume/Time |
|---|---|---|
| Acidic environment (pH 2) | 10 mM HCl | 3mL /overnight |
| Neutral environment (pH 7) | $H_2O$ | 3mL /7x30min |
| Dehydration series | 80:20 $H_2O$:EtOH | 3mL /15min |
| | 60:40 $H_2O$:EtOH | |
| | 40:60 $H_2O$:EtOH | |
| | 20:80 $H_2O$:EtOH | |
| | 100% EtOH | 3mL /15min (x4) |
| Rehydration | $H_2O$ | 3mL /30min (x2) |

## Spectroscopic characterisation of functionalised PVA products

**Table S2**. TB grafting measured in TB-functionalised (polyvinyl alcohol) (PVA-TB) and its methacrylated derivative (PVA-TB-MA) via UV spectroscopy of corresponding aqueous solutions.

| Sample ID | Abs (579 nm) | [TB] /mg $mL^{-1}$ | [Grafted TB] / mg·$g^{-1}$ |
|---|---|---|---|
| **PVA-TB** | 0.6412 | 0.0846 | 8.5±0.9 |
| **PVA-TB-MA** | 0.2813 | 0.0348 | 3.5±0.2 |

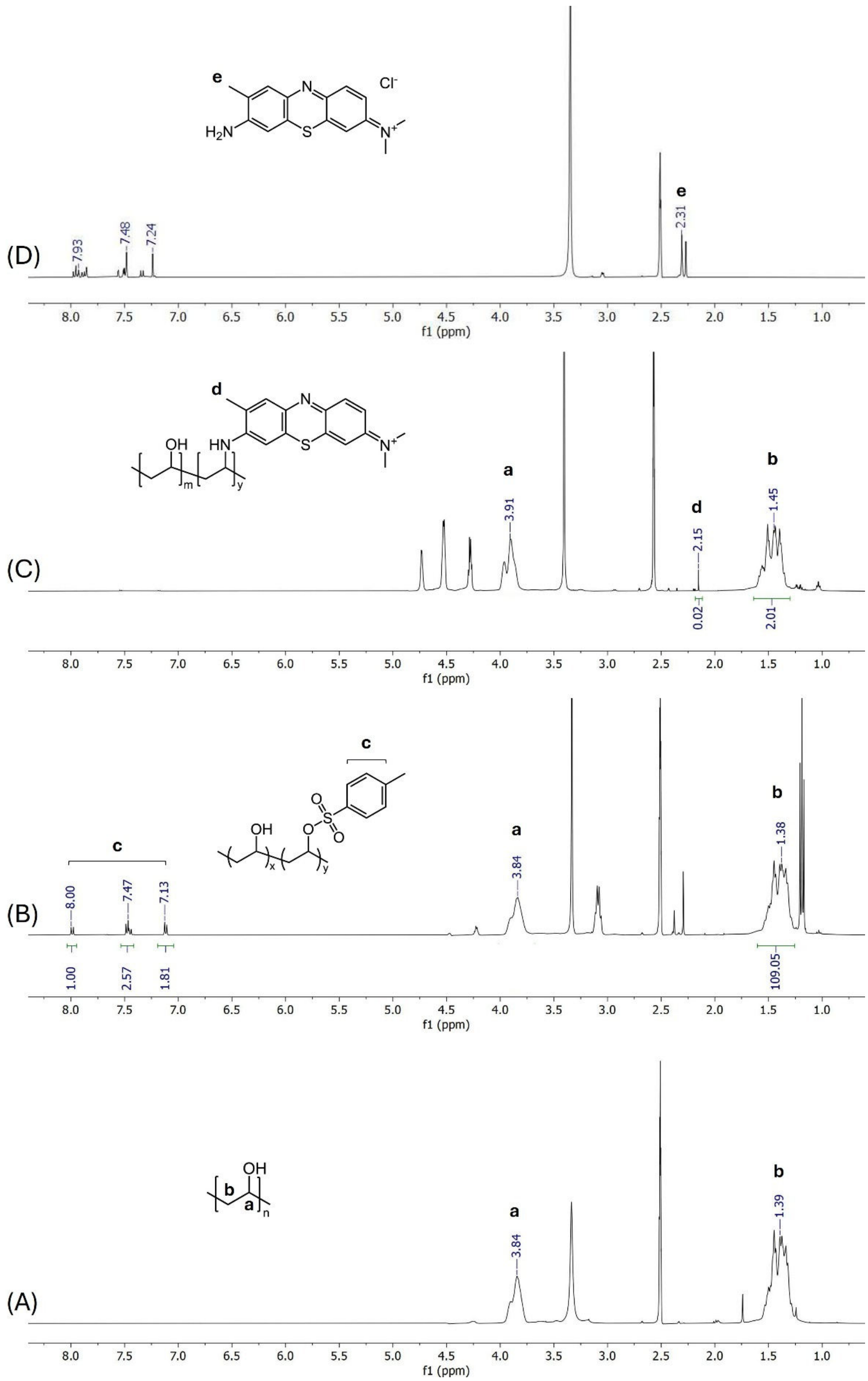

**Figure S3.** $^1$H NMR spectra recorded at room temperature in dimethyl sulfoxide-$d_6$ ($d_6$-DMSO). (A): PVA; (B): PVA-Ts; (C): PVA-TB; (D): TB. The appearance of aromatic tosyl resonances in PVA-Ts (B) and their disappearance following reaction with TB (C) support successful tosylation and subsequent substitution. A weak resonance at approximately 2.15 ppm was observed in PVA-TB (C) and is consistent with the methyl group of the phenothiazine residue (D). Owing to the low intensity of TB-associated signals, TB grafting was quantified independently by UV-Vis spectroscopy.

## Composition of UV-cured co-networks

**Table S3**. Chemical composition of UV-cured co-networks made of toluidine blue (TB)-grafted poly(vinyl alcohol) methacrylate (PVA-TB-MA) and type I collagen functionalised with 4-vinylbenzyl (4VB) residues.

| Sample ID | [Total Polymer] /wt.% | [MA]:[4VB] | Weight /mg | TB content /µg | [TB] /wt.% |
|---|---|---|---|---|---|
| P2VA14 | 2% | 14:1 | 17.7 | 31.5 | 0.18 |
| P2VA28 | 2% | 28:1 | 26.8 | 63 | 0.24 |
| P2VA32 | 2% | 32:1 | 20.8 | 50.4 | 0.24 |
| P2VA35 | 2% | 35:1 | 27.7 | 67.2 | 0.24 |
| P2VA39 | 2% | 39:1 | 29.6 | 77.7 | 0.26 |
| P2VA44 | 2% | 44:1 | 23.8 | 63 | 0.26 |
| P2VA47 | 2% | 47:1 | 28.8 | 84 | 0.27 |
| P2VA53 | 2% | 53:1 | 22.6 | 67.2 | 0.30 |
| P2VA88 | 2% | 88:1 | 30.9 | 94.5 | 0.31 |
| P3VA28 | 3% | 28:1 | 30.0 | 70 | 0.23 |
| P3VA35 | 3% | 35:1 | 30.0 | 75 | 0.25 |
| P3VA39 | 3% | 39:1 | 30.0 | 81 | 0.27 |

## TB release quantification

**Table S4**. TB leaching from samples of PHOVA, P3VA35 and 4VBC* following 96-hour incubation *in vitro* (1X PBS, pH 7.4) at either 25 or 37 °C. n.a.: Not applicable.

| Sample ID | Abs (592 nm) | TB (released) /µg | TB (grafted) /mg | TB retention /wt.% | Mass retention /wt.% |
|---|---|---|---|---|---|
| **PHOVA (25 °C)** | 0.0041 | 1.715±0.412 | 0.03 | 94.28±1.37 | 92.56±7.86 |
| **P3VA35 (25 °C)** | -0.0019 | 0.038±0.032 | 0.0237 | 99.84±0.14 | 98.99±2.74 |
| **4VBC* (25 °C)** | 0.0025 | n.a. | n.a. | n.a. | n.a. |
| **PHOVA (37 °C)** | 0.0018 | 1.074±0.028 | 0.0172 | 93.76±0.16 | 89.14±4.53 |
| **P3VA35 (37 °C)** | 0.0045 | 1.038±1.089 | 0.0271 | 93.23±4.00 | 98.63±1.18 |
| **4VBC* (37 °C)** | 0.0000 | n.a. | n.a. | n.a. | n.a. |

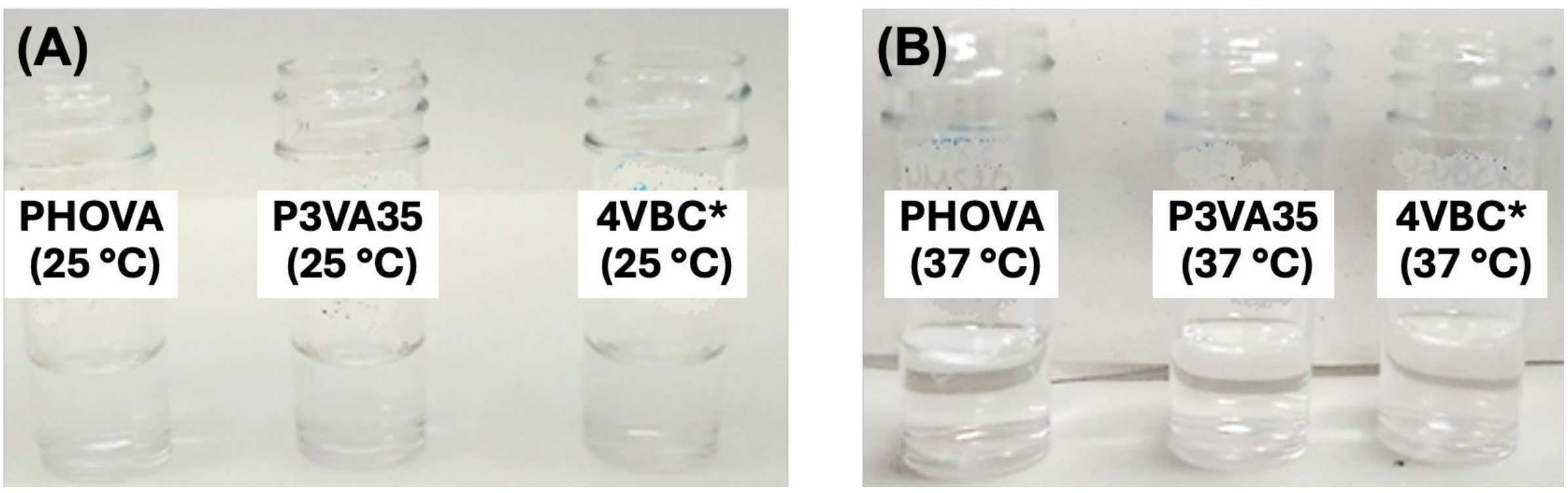

**Figure S4**. Appearance of the supernatant following 96-hour incubation (1X PBS, pH 7.4) of either UV-cured toluidine blue-grafted poly(vinyl alcohol) methacrylate (PHOVA), UV-cured co-network P3VA35 or UV-cured, 4-vinylbenzyl-functionalised type I collagen (4VBC*) at 25 °C (A) and 37 °C (B).

### Liquid chromatography-mass spectrometry (LC-MS)

LC-MS was carried out at room temperature on a Phenomenex Kinetex C18 column, 2.6 mm particle size, 50 × 2.1 mm I.D. column. Chromatography was carried out using two solvents: (A) water and 0.1% formic acid solution and (B) acetonitrile and 0.1% formic acid solution. A linear gradient programme was applied: 2–95% increase of solvent B over 1 min. The flow rate during the experiment was 1.3 mL·min$^{-1}$. Injections were made by a Basic Marathon autosampler equipped with a 20 mL loop. The method was carried out on a Thermo Ultimate 3000 UHPLC using a Bruker Amazon Speed IonTtrap for the MS detection and a Diode Array Detector. The ESI (electrospray ionisation) parameters in the positive ion mode were as follows: spray voltage 4000 V (applied to the spray tip needle), dry gas 10 dm$^3$·min$^{-1}$, dry temperature 365 °C, capillary 60 nA, nebulizer 65 psi ($N_2$).

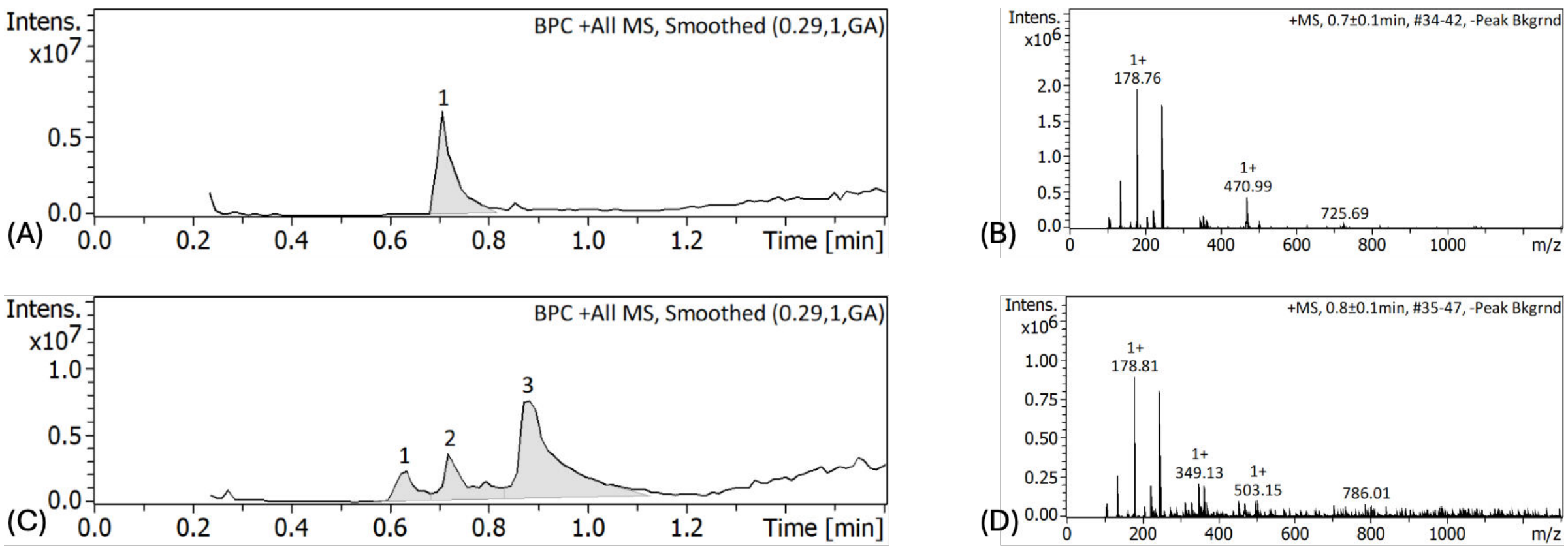

**Figure S5**. LC-MS spectrum of the photoinitiator (I2959) in 1X PBS (A) and the corresponding mass peaks (B), as well as LC-MS spectrum of the supernatant obtained via 72h-incubation (1X PBS, pH 7.4, 37 °C) of PHOVA (C), and the mass peaks corresponding to peak 2 (D), confirming the leaching of photoinitiator.

### UV-Vis spectroscopy during sample washing

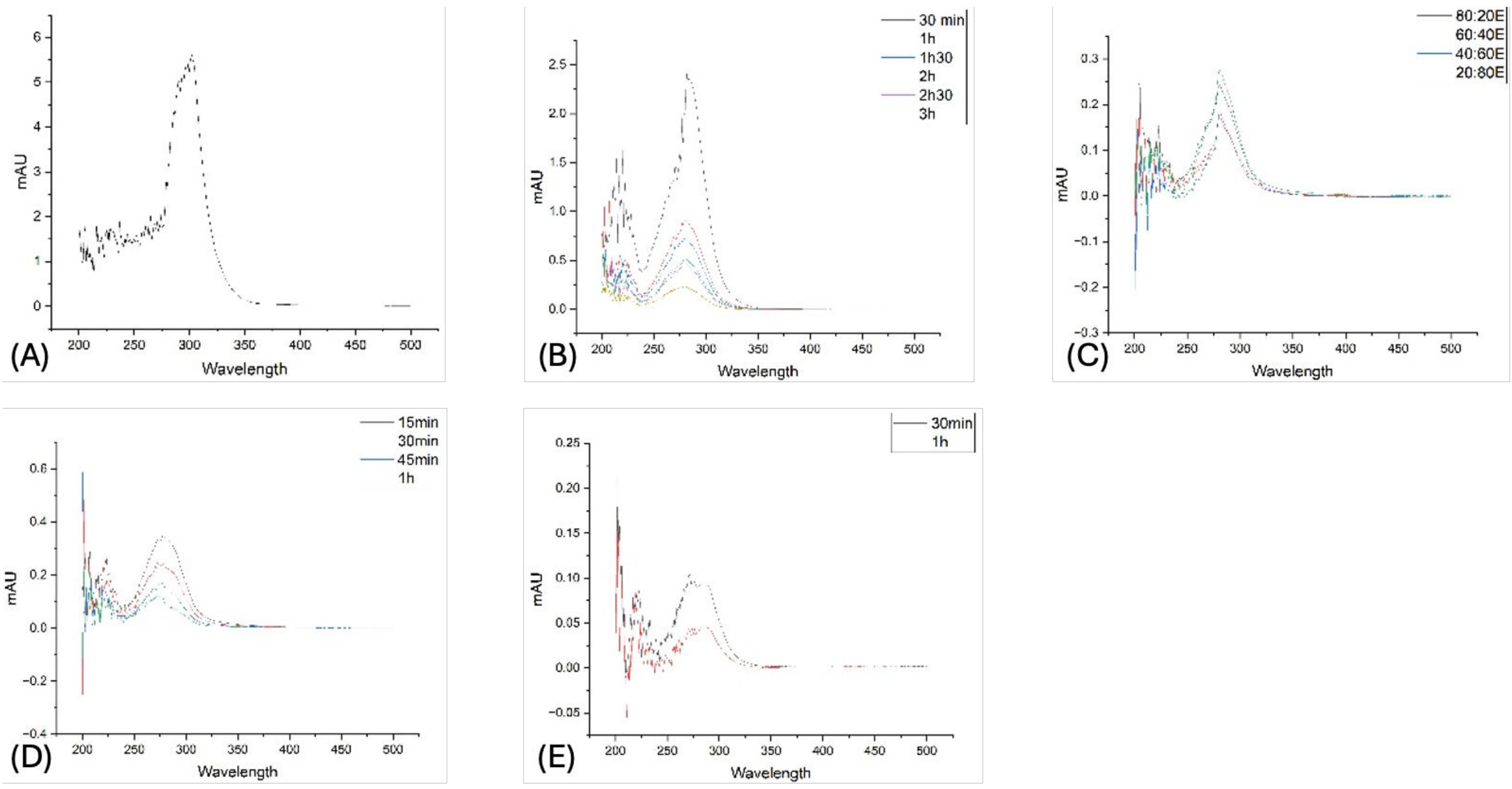

**Figure S6**. Absorption spectra of the supernatant recorded following incubation in 10 mM HCl (A), $H_2O$ (B), EtOH-$H_2O$ series (C), EtOH (D) and after rehydration after overnight drying (E). UV-Vis readings were collected at selected incubation time points in each medium.

## Characterisation of UV-cured interpenetrating network control

**Table S5**. Swelling ratio (SR), volumetric swelling (Q) and gel content (G) measured with sample I3VA35 after 24-hour incubation in either distilled water at 37 °C (SR and Q) or 10 mM HCl at room temperature (G).

| SR /wt.% | Q /vol% | G /wt.% |
|---|---|---|
| 482.09±28.19 | 297.8±52.53 | 72.2±4.41 |

**Table S6**. TB retention and mass retention measured via weight analysis and UV-Vis spectrophotometry, respectively, following 96-hour incubation (1X PBS, pH 7.4) of sample I3VA35 at either 25 or 37 °C.

| Incubation Temperature | Abs (592 nm) | TB (released) /μg | TB (grafted) /mg | TB retention /wt.% | Mass retention /wt.% |
|---|---|---|---|---|---|
| **25 °C** | 0.00196 | 3.92±0.57 | 0.0189 | 79.21±3.01 | 78.75±1.08 |
| **37 °C** | 0.00116 | 2.57±0.34 | 0.0229 | 88.76±1.48 | 82.44±5.16 |

## Wet-state tensile testing of UV-cured co-networks

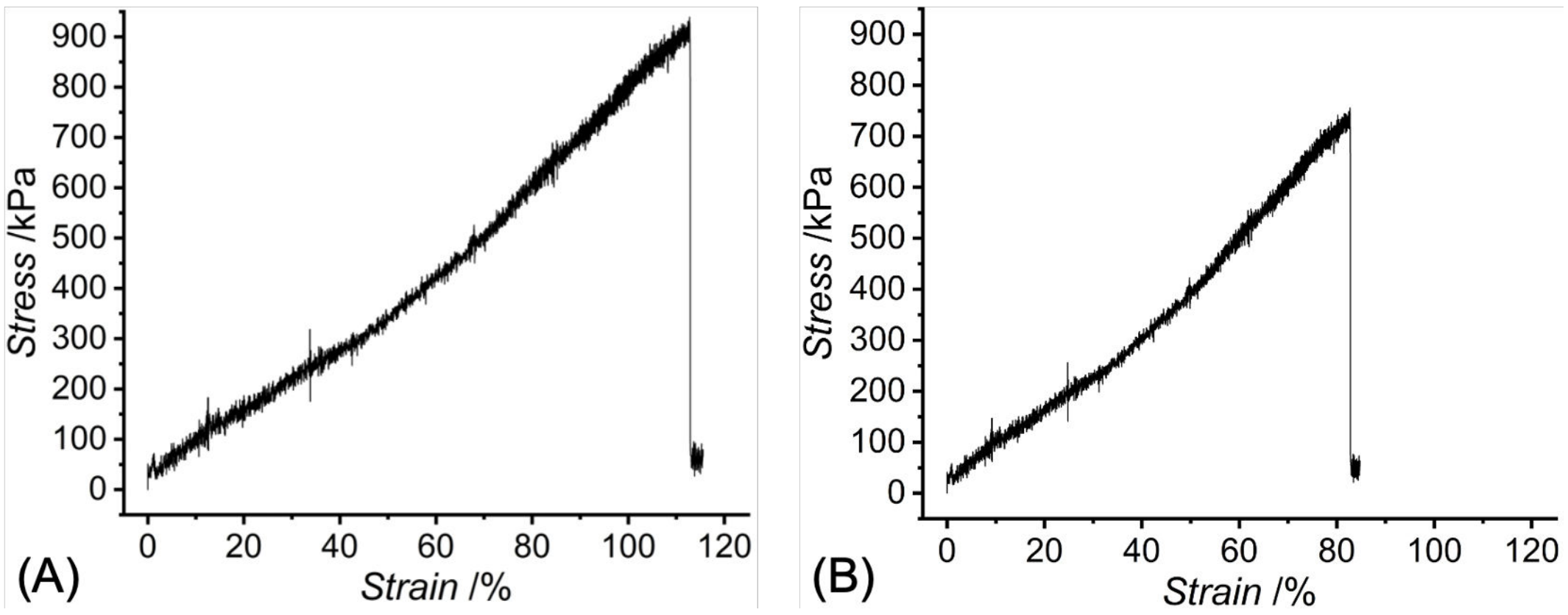

**Figure S7.** Representative stress-strain curves obtained via wet-state tensile testing of UV-cured co-networks P2VA39 (A) and P3VA39 (B).

## ROS assay

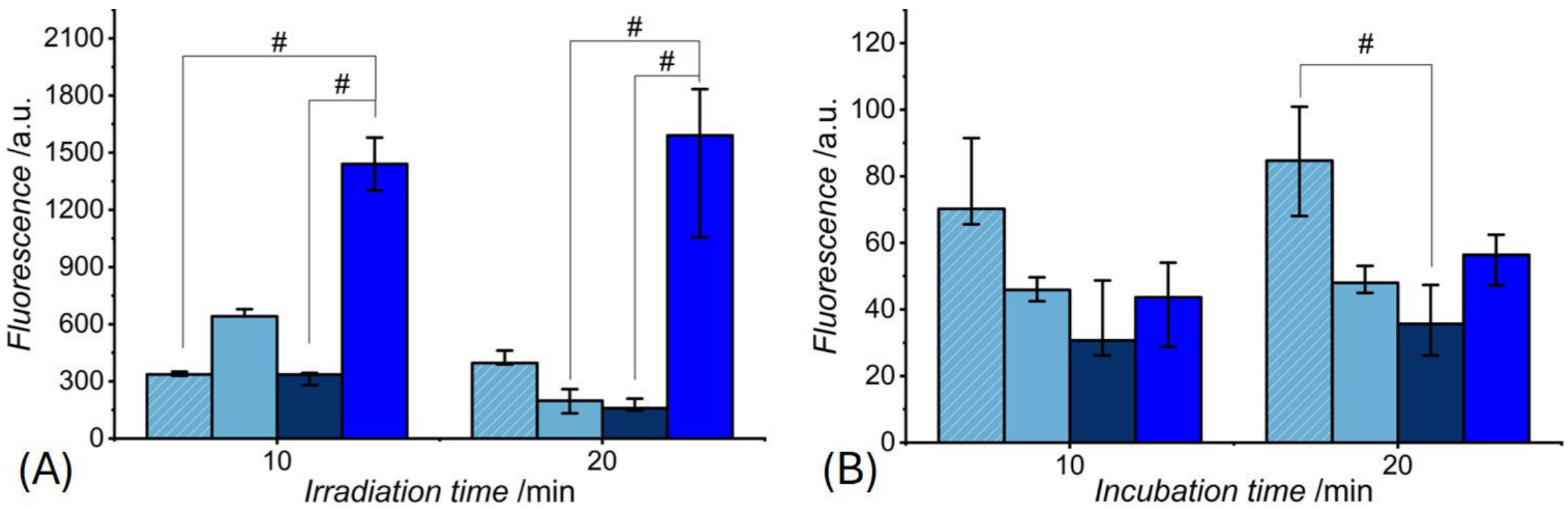

**Figure S8.** ROS-associated fluorescence determined via DCFH assay following either 10- and 20-min work light exposure (A) or 10- and 20-min dark incubation (B) of UV-cured samples. (▨): P3VA39; (■): P2VA39 (n=3); (■): P2VA88 (n=3); (■): PHOVA (n=2). Data are expressed as median values with error bars representing the distance from the median to first and third quartiles. Statistical significance was assessed using Kruskal–Wallis followed by Dunn's pairwise comparisons with Benjamini–Hochberg correction (#: $p < 0.05$).

## 2,2-diphenyl-1-picrylhydrazyl (DPPH) assay

**Table S7**. Radical Scavenging Activity (%) determined via DPPH assay in the presence of UV-cured 4-vinylbenzyl-functionalised type I collagen (4VBC*), UV-cured toluidine blue-grafted poly(vinyl alcohol) methacrylate (PHOVA) and corresponding UV-cured co-networks (P2VA39, P2VA88 and P3VA39).

| 4VBC* | PHOVA | P2VA39 | P2VA88 | P3VA39 |
|---|---|---|---|---|
| 24 | 43 | 39 | 49 | 36 |